\documentclass[twocolumn,aps,prb,10pt,nofootinbib]{revtex4-2}
\usepackage[utf8]{inputenc}
\usepackage[T1]{fontenc}
\usepackage{amsmath}
\usepackage{amssymb}
\usepackage{graphicx}
\usepackage{hyperref}
\usepackage{dsfont}
\usepackage{mathtools}
\usepackage[shortlabels]{enumitem}
\usepackage{physics}
\usepackage{tabularx}
\usepackage{booktabs}
\usepackage{bm}
\usepackage{tikz}

\hypersetup{
    colorlinks,
    linkcolor={blue},
    citecolor={blue},
    urlcolor={blue}
}

\graphicspath{{./}{./figs/}}

\newcommand{\ii}{\mathrm{i}}
\DeclareMathOperator{\diag}{diag}

\newcommand{\HH}{\mathcal{H}}
\newcommand{\hc}{\mathrm{H.c.}}
\newcommand{\prpr}{'\!'}
\newcommand{\SUtwocrossUone}{\mathrm{SU(2)}_{1,3} \times \mathrm{U(1)}}
\newcommand{\UonecrossUone}{\mathrm{U(1)} \times \mathrm{U(1)}}
\renewcommand{\vec}{\mathbf}

\begin{document}

%%%%%%%%%%%%%%%%%%%%%%%%%%%%%%%%%%%%%%%%%%%%%%%%%%%%%%%%%%%

\title{Fractionalization, emergent SU($N$) symmetries, and fragmentation\\in layered quantum spin-orbital models}

\author{Pedro M. C\^onsoli$^*$}
\author{Aayush Vijayvargia$^*$}
\author{Onur Erten}

\affiliation{Department of Physics, Arizona State University, Tempe, Arizona 85287, USA}
\date{\today}

%%%%%%%%%%%%%%%%%%%%%%%%%%%%%%%%%%%%%%%%%%%%%%%%%%%%%%%%%%%
\begin{abstract}

We propose a family of layered quantum spin-orbital models as a platform to study fractionalization, unconventional forms of symmetry-breaking order, and their possible coexistence. The models are built by stacking $N$ layers of a square-lattice system in which Kitaev-type interactions promote the formation of a $\mathbb{Z}_2$ quantum spin-orbital liquid and coupling the different layers via Ising spin interactions. Using a parton construction, we show how, at low energies, these Hamiltonians can be mapped to $N$-component Fermi Hubbard models on a $\pi$-flux square lattice at half filling. We also demonstrate that the models acquire an emergent SU($N$) symmetry in the limit of equal all-to-all interlayer couplings and argue that, for $N>2$, the proximity to this limit offers the potential to realize an array of competing phases. To illustrate this point, we compute the zero-temperature phase diagram of the effective $N=3$ Hubbard model within mean-field theory and uncover rich phenomena, including intertwined orders and flavor-selective localization. Mapping back to the original degrees of freedom reveals that the ground states realize distinct forms of magnetic fragmentation, wherein the orbitals remain in a quantum liquid state whereas the spins can present conventional long-range order or nonlocal order characterized by a nontrivial string order parameters. We highlight possible extensions of our construction as well as its potential to provide concrete microscopic models for different fractionalized quantum critical points.

\end{abstract}
%%%%%%%%%%%%%%%%%%%%%%%%%%%%%%%%%%%%%%%%%%%%%%%%%%%%%%%%%%%

\maketitle
\def\thefootnote{*}\footnotetext{These authors contributed equally to this work.}
\renewcommand{\thefootnote}{\arabic{footnote}}
\setcounter{footnote}{0}

%%%%%%%%%%%%%%%%%%%%%%%%%%%%%%%%%%%%%%%%%%%%%%%%%%%%%%%%%%%
%%%%%%%%%%%%%%%%%%%%%%%%%%%%%%%%%%%%%%%%%%%%%%%%%%%%%%%%%%%
%%%%%%%%%%%%%%%%%%%%%%%%%%%%%%%%%%%%%%%%%%%%%%%%%%%%%%%%%%%

\section{Introduction}
\label{sec:intro}

The past two decades have witnessed great strides in the use of synthetic platforms to simulate quantum many-body physics \cite{feynman82,georgescu14,blatt12,bloch12,guzik12,browaeys20}. In addition to providing access to states of matter -- namely, quantum spin liquids -- that have long eluded unambiguous observation in solids \cite{semeghini21,satzinger21}, this approach has boosted the exploration of phenomena either deemed unachievable or of unlikely occurrence in conventional materials. 
In particular, the ability to engineer systems of ultracold Fermi gases featuring SU($N$) symmetries with tunable $N \le 10$ \cite{rey14,gorshkov10,taie12,pagano14,zhang14,hofrichter16,taie22} has opened the door to realize a plethora of exotic phases predicted in theory. These range from novel forms of magnetic order \cite{toth10,corboz11,bauer12,consoli25} to nontrivial quantum disordered states  \cite{assaad05,hermele09,hermele11,nataf14}, which are favored at higher values of $N$ \cite{affleck88,marston89,arovas88}.

In traditional solid-state platforms, the emergence of SU($N$) symmetries with $N>2$ is frequently associated with Kugel-Khomskii models \cite{kugel82,li98}, i.e., effective low-energy Hamiltonians that describe the coupling between active spin and orbital degrees of freedom. 
Over the years, these models have been applied to various systems of strongly correlated electrons, including different types of Mott insulators \cite{tokura00,vernay04,natori16,yamada18,natori18,natori19}, iron-based superconductors \cite{krueger09}, and twisted bilayer graphene \cite{venderbros18,yuan18}.
However, their potential to realize exotic ground states \cite{corboz12,nussinov15,natori26} is limited by two aspects. The first is that, in realistic settings, physical mechanisms such as the Jahn-Teller effect \cite{jahn37} tend to spoil the orbital degeneracy required for an emergent SU($N$) symmetry. The second is that, in contrast to the cold-atom systems mentioned above, materials chemistry places strong restrictions on possible values of $N$. In fact, most known applications of Kugel-Khomskii models are for two-orbital systems, for which the relevant symmetry is SU(4).

In this paper, we present a family of tractable theoretical models with well defined SU($N$)-symmetric limits where the symmetry index $N$ is not tied to an internal degeneracy, but is rather structural and adjustable. Specifically, the models comprise $N$ identical layers of a quantum spin-orbital Hamiltonian featuring Kitaev-type, bond-dependent interactions on a square lattice \cite{Nakai_PRB2012,chulliparambil20,vijayvargia23}. 
When decoupled, these layers realize $N$ copies of an exact quantum spin-orbital liquid whose low-energy excitations are itinerant Majorana fermions coupled to a static $\mathbb{Z}_2$ gauge field \cite{kitaev06,Nakai_PRB2012}. 
A key design choice is made to ensure that, while the interlayer interactions spoil the exact solubility of our models, they conserve the intralayer fluxes. This guarantees that the systems under consideration are sufficiently close to the exactly-solvable limit to admit a controlled low-energy description in terms of $N$-flavor complex fermions. With this, we set out to demonstrate that these models can harbor a wealth of nontrivial phenomena, especially in the vicinity of their SU($N$)-symmetric limits.

The remainder of this paper is organized as follows. In Sec.~\ref{sec:model}, we introduce the aforementioned family of layered spin-orbital models and show that, at low energies, they can be mapped onto Fermi Hubbard models with $N$-flavor fermions hopping on a $\pi$-flux square lattice at half filling. We then discuss the relevant symmetries of these effective models, placing special emphasis on emergent SU($N$) symmetries.
Subsequently, in Sec.~\ref{sec:mft_general}, we develop a mean-field framework to describe the zero-temperature phases realized in the Hubbard models with sufficiently small $N$. After highlighting conceptual differences between the $N=2$ case and its $N>2$ counterparts, we provide a detailed analysis of the $N=3$ theory in Sec.~\ref{sec:mft}. In particular, we solve the mean-field equations for a restricted form of interlayer interactions and uncover a rich phase diagram with different types of symmetry-breaking order, some of which are found to be intertwined \cite{fradkin15}.
Section~\ref{sec:mapback} proceeds to translate these results back to the original spin and orbital degrees of freedom. Finally, Sec.~\ref{sec:summary} closes the paper with a summary of our results and an outlook.
The appendices provide details on our mean-field calculations as well as proofs pertaining to the structure of the SU(3) group and the development of nontrivial string correlations.

%%%%%%%%%%%%%%%%%%%%%%%%%%%%%%%%%%%%%%%%%%%%%%%%%%%%%%%%%%%
%%%%%%%%%%%%%%%%%%%%%%%%%%%%%%%%%%%%%%%%%%%%%%%%%%%%%%%%%%%
%%%%%%%%%%%%%%%%%%%%%%%%%%%%%%%%%%%%%%%%%%%%%%%%%%%%%%%%%%%

\section{Spin-orbital models and their relation to $N$-flavor fermionic Hubbard models}
\label{sec:model}

\begin{figure}[t!]
    \centering
    \includegraphics[width=\linewidth]{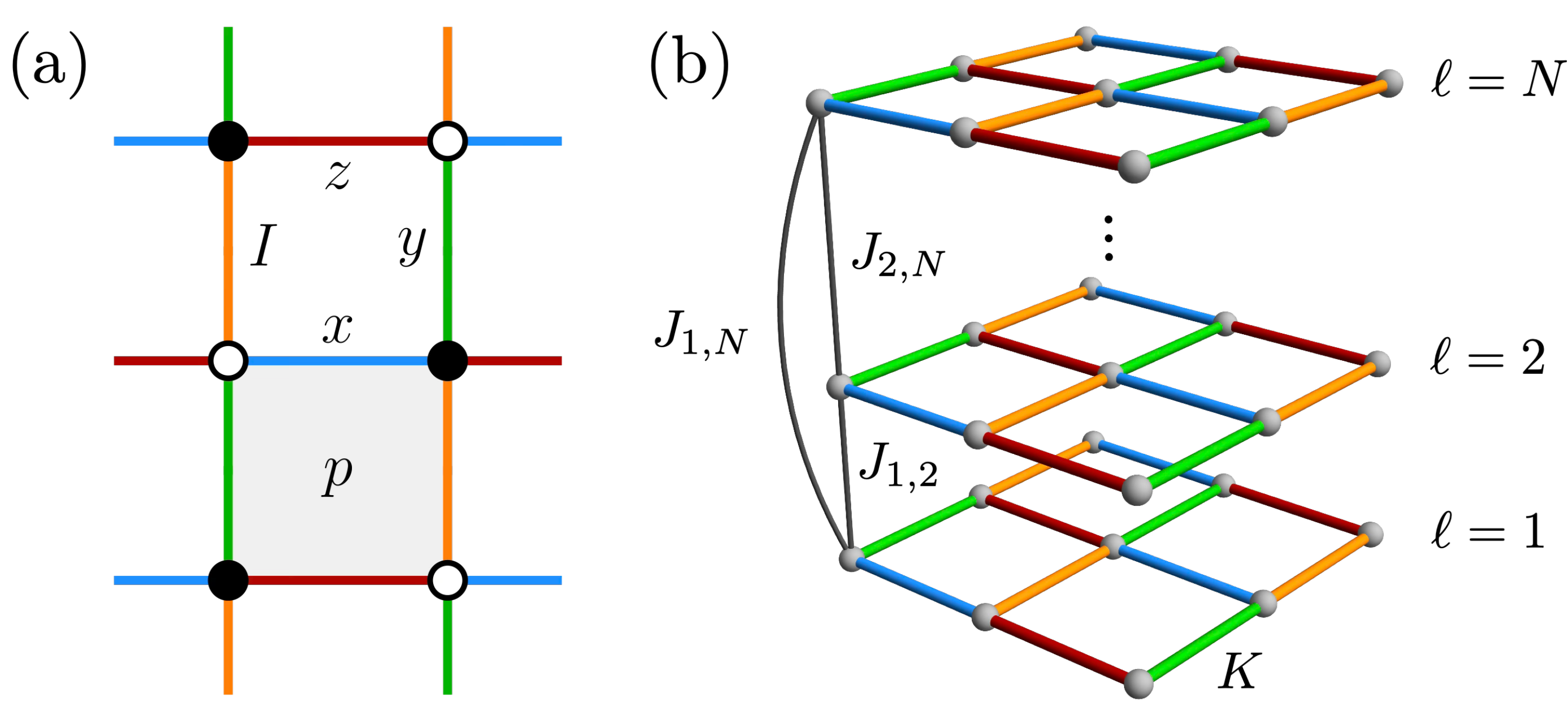}
    \caption{Schematics illustrating the spin-orbital model in Eqs.~\eqref{eq:Hso}-\eqref{eq:HsoJ}. (a) Section of a single square layer, showing the four different types of bonds, labeled $x$, $y$, $z$, and $I$, and highlighting one of the two inequivalent types of elementary square plaquettes. The two crystallographic sublattices, $A$ and $B$, are represented by black and white sites, respectively. (b) Multilayer system of $N$ identical layers stacked without any relative displacement.}
    \label{fig:somodel}
\end{figure}

We consider a class of Kitaev-type spin-orbital models \cite{Saptarshi_2009, Tikhonov_PRL_2010, YaoLee2011, Chua_PRB_2011, Carvalho2018, Akram_Vison_2023, Nica2023_NPJ, Keskiner_PRB_2023, Poliakov2024, KESKINER2025, Vijayvargia_PRL2025, sobral25, Akram_npj2026} in which spin ($\sigma$) and orbital ($\tau$) degrees of freedom are distributed over the sites of $N$ perfectly stacked layers of a square lattice. The Hamiltonian
\begin{equation}
    \HH = \HH_K + \HH_J
    \label{eq:Hso}
\end{equation}
is composed of a sum of two terms,
\begin{align}
	\HH_K &= -K \sum_{\ell=1}^{N} \sum_{\alpha} \sum_{\langle ij \rangle_\alpha} 
	\left( \sigma_{i\ell}^x \sigma_{j\ell}^x + \sigma_{i\ell}^y \sigma_{j\ell}^y \right) \otimes \tau_{i\ell}^\alpha \tau_{j\ell}^\alpha,
    \label{eq:HsoK}
	\\
	\HH_J &= \sum_i \sum_{\ell' > \ell} J_{\ell\ell'} \sigma_{i\ell}^z \sigma_{i\ell'}^z,
	\label{eq:HsoJ}
\end{align}
where the index $i$ specifies the position of a site within a given layer $\ell = 1,\ldots,N$ and $\alpha \in \left\{x,y,z,I\right\}$ labels the four types of in-plane bonds (see Fig.~\ref{fig:somodel}). The symbols $\sigma^\alpha$ and $\tau^\alpha$ with $\alpha = x,y,z$ denote Pauli matrices, and $\tau^I$ stands for the $2\times 2$ identity matrix. Thus, $\HH_K$ describes nearest-neighbor intralayer interactions of XY and Kitaev type for spins and orbitals, respectively, whereas $\HH_J$ couples spins with the same in-plane coordinate $i$ via an Ising interaction. The coupling constants $J_{\ell\ell'} = J_{\ell'\ell}$ between layers $\ell$ and $\ell'$ are in principle arbitrary. However, we will henceforth assume that they are all antiferromagnetic, $J_{\ell\ell'} > 0$, for concreteness.

In the absence of interlayer couplings ($J_{\ell\ell'} = 0$), the Hamiltonian $\HH$ reduces to $N$ copies of the spin-orbital model proposed in Ref.~\onlinecite{Nakai_PRB2012}. Similarly to the Kitaev honeycomb model \cite{kitaev06}, each such copy features an extensive number of independent conserved quantities $\hat{W}_p$ corresponding to the fluxes through elementary square plaquettes. If $i$, $j$, $m$, and $n$ denote the four sites in a plaquette $p$ (we omit the layer index here for brevity), ordered clockwise and such that $i$ and $j$ are connected by an $I$ bond [see Fig.~\ref{fig:somodel}(a)], then we can define\footnote{This definition yields the correct operators for the two types of inequivalent plaquettes in the system \cite{chulliparambil20}.}
\begin{equation}
    \hat{W}_p = \sigma_i^z \sigma_j^z \tau_i^x \tau_j^y \tau_m^x \tau_n^y.
    \label{eq:Wp}
\end{equation}
Given that different plaquette operators also commute with each other, the Hamiltonian $\HH_K$ can be decomposed into sectors defined by the set of eigenvalues $W_p = \pm1$ of $\hat{W}_p$ for every $p$. This feature ensures that $\HH_K$ can be solved exactly by employing the Majorana fermion representation presented in Sec.~\ref{subsec:mapping}. In doing so, one finds that the ground state of $\HH_K$ is given by $N$ copies of a $\mathbb{Z}_2$ quantum spin-orbital liquid whose low-energy excitations are itinerant Majorana fermions moving on a static flux background.% with $W_p = -1$ on every plaquette, as dictated by Lieb's theorem \cite{lieb94}.

The inclusion of interlayer couplings $J_{\ell\ell'}$ perturbs the ground state of $\HH_K$ by promoting correlations between spin degrees of freedom $\bm{\sigma}_{i\ell}$ with fixed $i$. Furthermore, it spoils the exact solvability of the model since $\HH_J$ does not conserve fluxes through plaquettes that connect different layers $\ell$.
However, the \emph{intralayer} fluxes $\hat{W}_{p\ell}$ (we now restore the layer index for clarity) remain constants of motion \cite{vijayvargia23}. In the next subsection, we show how this feature assists in the investigation of our general $N$-layer models after $\HH$ is mapped onto a two-dimensional, $N$-flavor Fermi-Hubbard model where the hopping between nearest-neighbor sites is mediated by in-plane gauge fields.

%%%%%%%%%%%%%%%%%%%%%%%%%%%%%%%%%%%%%%%%%%%%%%%%%%%%%%%%%%%
%%%%%%%%%%%%%%%%%%%%%%%%%%%%%%%%%%%%%%%%%%%%%%%%%%%%%%%%%%%

\subsection{Mapping to $N$-flavor Hubbard models}
\label{subsec:mapping}

The mapping of Eqs.~\eqref{eq:Hso}-\eqref{eq:HsoJ} to an $N$-flavor fermionic Hubbard model is achieved in three steps \cite{chulliparambil21, vijayvargia23}. The first is to recast the Hamiltonian in terms of $4\times4$ Dirac matrices that span the full Hilbert space (including spin and orbital degrees of freedom) at each site. Omitting site indices for the moment, we define
\begin{align}
	\Gamma^1 &= -\sigma^y \otimes \tau^x,
	&
	\Gamma^2 &= -\sigma^y \otimes \tau^y,
	&
	\Gamma^3 &= -\sigma^y \otimes \tau^z,
	\notag \\
	\Gamma^4 &= \sigma^x \otimes \tau^I,
	&
	\Gamma^5 &= -\sigma^z \otimes \tau^I.
    \label{eq:Gamma mats}
\end{align}
These matrices satisfy the Clifford algebra $\left\{ \Gamma^\alpha, \Gamma^\beta \right\} = 2\delta_{\alpha\beta}$ and can be used to define the complementary set of matrices $\Gamma^{\alpha\beta} = \frac{\ii}{2} \comm{\Gamma^\alpha}{\Gamma^\beta}$. With this, Eqs.~\eqref{eq:HsoK} and \eqref{eq:HsoJ} can be rewritten as
\begin{align}
	\HH_K &= -K \sum_\ell \sum_\alpha \sum_{\langle ij \rangle_\alpha} \left( \Gamma_{i\ell}^\alpha \Gamma_{j\ell}^\alpha + \Gamma_{i\ell}^{\alpha5} \Gamma_{j\ell}^{\alpha5} \right),
    \label{eq:HGammaK}
	\\
	\HH_J &= \sum_{i, \ell} \sum_{\ell' > \ell} J_{\ell\ell'} \Gamma_{i\ell}^5 \Gamma_{i\ell'}^5,
    \label{eq:HGammaJ}
\end{align}
if one identifies the bond indices $(x,y,z,I) \equiv (1,2,3,4)$.

The second step is to represent the above $\Gamma$ matrices in terms of six Majorana fermions, $b^\mu$ ($\mu=1,\cdots,5$) and $c$, obeying $\left\{b^\mu, b^\nu\right\} = 2\delta_{\mu\nu}$, $\left\{ b^\mu, c \right\} = 0$, and $c^2 = 1$. The representation \cite{yao09,wu09,ryu09}
\begin{align}
	\Gamma^\mu &= \ii b^\mu c,
	&
	\Gamma^{\mu\nu} &= \ii b^\mu b^\nu,
	\label{eq:Gamma rep}
\end{align}
is then faithful in the subspace defined by the constraint
\begin{equation}
	D = \ii b^1 b^2 b^3 b^4 b^5 c = 1.
	\label{eq:majorana constraint}
\end{equation}
By using Eq.~\eqref{eq:Gamma rep} and relabeling $(b^5,c) \mapsto (c^x,c^y)$, one finds the following Majorana representation of Eqs.~\eqref{eq:HGammaK} and \eqref{eq:HGammaJ}:
\begin{align}
	\tilde{\HH}_K &= K \sum_\ell \sum_\alpha \sum_{\langle ij \rangle_\alpha} \ii \hat{u}_{\ell, ij}^\alpha \left( c_{i\ell}^x c_{j\ell}^x + c_{i\ell}^y c_{j\ell}^y \right),
    \label{eq:HMajK}
	\\
	\tilde{\HH}_J &= - \sum_{i, \ell} \sum_{\ell' > \ell} J_{\ell\ell'} c_{i\ell}^x c_{i\ell}^y c_{i\ell'}^x c_{i\ell'}^y,
    \label{eq:HMajJ}
\end{align}
where $\hat{u}_{\ell,ij}^\alpha = \ii b_{i\ell}^\alpha b_{j\ell}^\alpha$ are operators defined on the links of layer $\ell$ with the convention that $i$ and $j$ belong to sublattices $A$ and $B$ [see Fig.~\ref{fig:somodel}(a)], respectively.
The tildes in Eqs.~\eqref{eq:HMajK} and \eqref{eq:HMajJ} serve as reminders that the Hamiltonian now acts on an enlarged Hilbert space, and that eigenstates of $\tilde{\HH} = \tilde{\HH}_K + \tilde{\HH}_{J}$ are in general unphysical. As in Ref.~\onlinecite{kitaev06}, one can render an unphysical eigenstate physical by acting on it with the projector
\begin{equation}
    P = \prod_{i\ell} \left( \frac{1+D_{i\ell}}{2} \right),
    \label{eq:projector}
\end{equation}
which imposes the constraint in Eq.~\eqref{eq:majorana constraint} on every site.

As noted above, although the interaction $\tilde{\HH}_J$ spoils the integrability of the Hamiltonian, it preserves the property that all $\hat{u}_{\ell,ij}^\alpha$ operators are conserved, i.e., 
\begin{equation}
    \comm{\hat{u}_{\ell,ij}^\alpha}{\tilde{\HH}} = \comm{\hat{u}_{\ell,ij}^\alpha}{\hat{u}_{\ell',i'j'}^{\alpha'}} = 0.
\end{equation}
Hence, as in Kitaev's honeycomb model \cite{kitaev06}, $\tilde{\HH}$ can be decomposed into sectors $\left\{u\right\}$ defined by different sets of eigenvalues $u_{\ell,ij}^\alpha = \pm 1$ of $\hat{u}_{\ell,ij}^\alpha$.

The third and final step required to complete the mapping to an $N$-flavor Hubbard model is to combine the $c^x$ and $c^y$ itinerant Majorana fermions at each site into a single complex fermion $f$:
\begin{align}
	c_{i\ell}^x &= f_{i\ell}^\dagger + f_{i\ell}, 
	&
	c_{i\ell}^y &= \frac{f_{i\ell}^\dagger - f_{i\ell}}{\ii}.
	\label{eq:ctof}
\end{align}
This allows us to finally write the Hamiltonian as
\begin{align}
	\tilde{\HH}_K &= 2K \sum_{\ell=1}^N \sum_{\alpha} \sum_{\langle ij \rangle_\alpha} \hat{u}_{\ell, ij}^\alpha \left( \ii f_{i\ell}^\dagger f_{j\ell} + \hc \right),
    \label{eq:HhubbK}
    \\
    \tilde{\HH}_J &= 4 \sum_{i, \ell} \sum_{\ell' > \ell} J_{\ell\ell'} \left( \hat{n}_{i\ell} - \frac{1}{2} \right) \left( \hat{n}_{i\ell'} - \frac{1}{2} \right),
	\label{eq:HhubbJ}
\end{align}
where $\hat{n}_{i\ell} =  f_{i\ell}^\dagger f_{i\ell}$. Within a fixed gauge sector $\left\{ u \right\}$, $\tilde{\HH}_K$ can be interpreted as a hopping term for complex fermions $f_{i\ell}$ of flavors $\ell = 1,\ldots,N$ on a square lattice with sites $i$. While the hopping amplitudes have a fixed absolute value of $2K$, their signs are controlled by the $\mathbb{Z}_2$ gauge fields $u_{\ell,ij}^\alpha$. The interacting term $\tilde{\HH}_J$, on the other hand, describes repulsive on-site interactions, since we assume that $J_{\ell\ell'} > 0$.

Lieb's theorem \cite{lieb94,macris96} asserts that, regardless of the values of the 
couplings $J_{\ell\ell'}$, the ground states of $\tilde{\HH}$ are realized for configurations $\{ u \}$ that yield a $\pi$ flux on every elementary intralayer plaquette\footnote{Based on Lieb's original proof \cite{lieb94}, we note that the applicability of the theorem does not require SU($N$) symmetry, but only: (i) bipartiteness of the lattice, (ii) particle-hole symmetry and reflection positivity of interactions, and (iii) mirror symmetries of the absolute values of hopping amplitudes. All of these conditions are satisfied by our models.}. Given that the Majorana representation of Eq.~\eqref{eq:Wp} is $\hat{W}_{p\ell} = \prod_{(ij) \in p} \hat{u}_{\ell,ij}^\alpha$, this condition can be implemented, for instance, by fixing $u_{\ell, ij}^x = -1$ and $u_{\ell, ij}^\alpha = +1$ for $\alpha \in \left\{ y,z,I \right\}$.
In this optimal flux sector, the noninteracting term $\tilde{\HH}_K$ of the Hubbard model can be diagonalized straightforwardly via a Fourier transformation. Its spectrum is composed of $2N$ flavor-degenerate bands
\begin{equation}
    \varepsilon_{\vec{k}\ell\pm} = \pm 4K\sqrt{ \sin^2 k_x + \cos^2 k_y },
    \label{eq:HK_spectrum}
\end{equation}
which display two independent Dirac cones at momenta $\vec{K} = (0,\pm\pi/2)$. Consequently, the system has a vanishing density of states at the Fermi level and should remain in a Dirac semimetallic phase for at least $J_{\ell\ell'} \ll K$.

%%%%%%%%%%%%%%%%%%%%%%%%%%%%%%%%%%%%%%%%%%%%%%%%%%%%%%%%%%%
%%%%%%%%%%%%%%%%%%%%%%%%%%%%%%%%%%%%%%%%%%%%%%%%%%%%%%%%%%%

\subsection{Gauge redundancy and symmetries}
\label{subsec:symm}

We now turn to a description of the gauge redundancy and of certain symmetries realized in the above family of spin-orbital models. This will be important for understanding the phase diagram reported in Sec.~\ref{sec:mft} and, more broadly, for appreciating the fact that $N>2$ can lead to novel phenomena.

%%%%%%%%%%%%%%%%%%%%%%%%%%%%%%%%%%%%%%%%%%%%%%%%%%%%%%%%%%%
\subsubsection{Gauge redundancy}

The parton representation in Eq.~\eqref{eq:Gamma rep} introduces a redundancy under local $\mathbb{Z}_2$ gauge transformations
\begin{align}
	\left( b_{i\ell}^\alpha, c_{i\ell}^x, c_{i\ell}^y \right)
	\longmapsto
	-\left( b_{i\ell}^\alpha, c_{i\ell}^x, c_{i\ell}^y \right),
	\label{eq:majorana gauge}
\end{align}
which preserve Eqs.~\eqref{eq:Gamma rep} and \eqref{eq:majorana constraint} as well as any physical observable, including the intralayer fluxes $\hat{W}_{p\ell}$ in Eq.~\eqref{eq:Wp}. Due to this redundancy, when fixing a flux sector $\left\{ W_{p\ell }\right\}$, one can always choose a gauge $\left\{ u \right\}$ where the link variables show no layer dependence, i.e., $u_{\ell,ij}^\alpha = u_{ij}^\alpha$.

By combining Eqs.~\eqref{eq:ctof} and \eqref{eq:majorana gauge}, we find that the gauge transformations above act on the complex fermions according to:
\begin{equation}
	\left( f_{i\ell}, f_{i\ell}^\dagger \right) 
	\longmapsto 
	-\left( f_{i\ell}, f_{i\ell}^\dagger \right).
	\label{eq:fermion gauge}
\end{equation}
In Sec.~\ref{sec:mapback}, we will see that the $\mathbb{Z}_2$ structure of these transformations has profound implications for the nature of the ground states of the system.

%%%%%%%%%%%%%%%%%%%%%%%%%%%%%%%%%%%%%%%%%%%%%%%%%%%%%%%%%%%
\subsubsection{Symmetries}

From the original form of the Hamiltonian, Eqs.~\eqref{eq:HsoK} and \eqref{eq:HsoJ}, one can easily verify that each layer $\ell$ separately conserves its spin magnetization along the $z$ axis, since all $N$ operators $M_\ell^z = \sum_i \sigma_{i\ell}^z$ commute with $\HH$. Using the identity $\sigma_{i\ell}^z = 1-2f_{i\ell}^\dagger f_{i\ell}$, this translates to the conservation of the total number of fermions of flavor $\ell$, which is tied to the invariance of $\tilde{\HH}$ under layer-dependent U(1) transformations
\begin{align}
    U(1)_\ell: 
    \quad
    f_{i\ell}^\dagger &\longmapsto e^{\ii \theta_\ell} f_{i\ell}^\dagger,
    &
    f_{i\ell} \longmapsto e^{-\ii \theta_\ell} f_{i\ell}.
    \label{eq:U1l}
\end{align}
These transformations can be generated by $M_\ell^z$ or, more directly, by the number operators $Q_\ell = \sum_i f_{i\ell}^\dagger f_{i\ell}$.

The use of the Majorana representation of the Dirac matrices is helpful in revealing further symmetries. Specifically, it is easy to verify that Eqs.~\eqref{eq:HhubbK} and \eqref{eq:HhubbJ} are invariant under the particle-hole transformation\footnote{Note that this transformation does not include a sublattice-dependent sign $(-1)^i$ that often appears in descriptions of Hubbard models on bipartite lattices (see, e.g., Ref.~\onlinecite{arovas22}). The lack of such a sign here is due to the presence of the imaginary unit $\ii$ in Eq.~\eqref{eq:HhubbK}.}
\begin{align}
    \mathrm{PH}: 
    \quad
    (f_{i\ell}^\dagger,f_{i\ell}) &\longmapsto 
    \Xi \, (f_{i\ell}^\dagger,f_{i\ell}) \, \Xi^{-1} = (f_{i\ell},f_{i\ell}^\dagger)
    \label{eq:ph}
\end{align}
implemented by
\begin{equation}
    \Xi = \prod_{j\ell} \left(\ii b_{j\ell}^1 c_{j\ell}^y \right)
    = \prod_{j\ell} \Gamma_{j\ell}^1.
\end{equation}
As noted above, the presence of this symmetry is crucial for the applicability of Lieb's theorem. Furthermore, within a fixed gauge sector $\{u\}$, the Hubbard model given by Eqs.~\eqref{eq:HhubbK} and \eqref{eq:HhubbJ} remains strictly at half filling unless this particle-hole symmetry is broken spontaneously -- a possibility we shall neglect in this paper.

In the special case where the system features identical all-to-all interlayer interactions, $J_{\ell\ell'} = J$, the Hamiltonian $\tilde{\HH}$ also displays an \emph{emergent} SU($N$) symmetry within any fixed sector $\left\{ u \right\}$. This follows from the invariance of Eqs.~\eqref{eq:HhubbK} and \eqref{eq:HhubbJ} under
\begin{align}
    \mathrm{SU}(N):
    \quad
    f_{i\ell}^\dagger &\longmapsto \sum_{\ell'} \mathcal{U}_{\ell\ell'} f_{i\ell'}^\dagger,
    \notag \\
    f_{i\ell} &\longmapsto  \sum_{\ell'}f_{i\ell'} \left( \mathcal{U}^\dagger \right)_{\ell'\ell},
    \label{eq:SUN}
\end{align}
where $\mathcal{U}$ is a unitary $N\times N$ matrix satisfying $\det \mathcal{U} = 1$. The transformations in Eq.~\eqref{eq:SUN} are generated by a set of $(N^2-1)$ operators
\begin{equation}
    S^a = 
    \sum_i S_i^a =
    \frac{1}{2} \sum_i \sum_{\ell,\ell'} f_{i\ell}^\dagger \eta_{\ell\ell'}^a f_{i\ell'}
    \label{eq:SUN generators}
\end{equation}
defined in terms of the Hermitian and traceless generators $\eta^a$ of the fundamental representation of SU($N$). Here, we choose to normalize these $N\times N$ matrices according to $\Tr \left( \eta^a \eta^b \right) = 2\delta_{ab}$, so that they correspond to the usual Pauli matrices $\sigma^a$ for $N=2$ and the Gell-Mann matrices $\lambda^a$ for $N=3$.

We emphasize that the Hamiltonian $\tilde{\HH}$ is \emph{never} truly SU($N$) symmetric, because it does not commute with the full set of generators in Eq.~\eqref{eq:SUN generators}. However, the aforementioned SU($N$) symmetry emerges for $J_{\ell\ell'} = J$ whenever the $\mathbb{Z}_2$ gauge fields remain static, i.e., the $\hat{u}_{ij,\ell}^\alpha$ operators can be replaced by integers $u_{ij,\ell}^\alpha$ reproducing a certain flux configuration. Given the gapped nature of the vison excitations, this condition is in fact met at temperatures much smaller than the vison gap. Hence, it is meaningful to speak of potential SU($N$) symmetry breaking in analyzing the ground-state properties of $\tilde{\HH}$. The breaking of this symmetry can be detected by tracking the local ``magnetizations''
\begin{equation}
    m_i^a = \expval{ S_i^a },
    \label{eq:mia}
\end{equation}
where $\expval{\cdots}$ denotes a ground-state expectation value.

%%%%%%%%%%%%%%%%%%%%%%%%%%%%%%%%%%%%%%%%%%%%%%%%%%%%%%%%%%%
%%%%%%%%%%%%%%%%%%%%%%%%%%%%%%%%%%%%%%%%%%%%%%%%%%%%%%%%%%%
%%%%%%%%%%%%%%%%%%%%%%%%%%%%%%%%%%%%%%%%%%%%%%%%%%%%%%%%%%%

\section{Mean-field theories for small $N$: General considerations}
\label{sec:mft_general}

In this section, we develop a mean-field treatment of the $N$-flavor Hubbard model $\tilde{\HH} = \tilde{\HH}_K + \tilde{\HH}_J$ defined by Eqs.~\eqref{eq:HhubbK} and \eqref{eq:HhubbJ}. 
To select a reasonable decoupling scheme for the interaction term $\tilde\HH_J$, we base ourselves on earlier studies of repulsive SU$(N)$ Hubbard models on the \emph{zero-flux} square lattice at half filling \cite{honerkamp04,miyatake10}. In particular, the renormalization-group analysis in Ref.~\onlinecite{honerkamp04} showed that, when $N\le 6$, the metallic phase realized in the noninteracting limit of the said Hubbard models has a weak-coupling instability toward a flavor density wave (FDW) with a $\vec Q=(\pi,\pi)$ ordering wavevector. Moreover, an explicit mean-field calculation for $N=3$ revealed that the FDW order can coexist with a charge density wave (CDW) \cite{honerkamp04}.
Motivated by these results, we assume that the $\pi$-flux systems considered here are prone to the same type of instability for sufficiently small $N$ and decouple $\tilde\HH_J$ in the density-density and particle-hole channels. This yields $N^2$ mean fields $O_{i}^{\ell\ell'} =  \expval{f_{i\ell}^\dagger f_{i\ell'}}$ per site, in terms of which the interaction term becomes
\begin{align}
    \tilde{\HH}_{J,\mathrm{MF}} &=
    \sum_i \sum_{\ell=1}^{N} \sum_{\ell\ne \ell'} J_{\ell\ell'} \left[
    2\left( \abs{O_i^{\ell\ell'}}^2 - O_i^{\ell\ell}O_i^{\ell'\ell'} \right)
    \right.
    \notag \\
    &\left.
    +4\left( O_i^{\ell'\ell'} - \frac{1}{2} \right) \hat{n}_{i\ell}
    -4O_i^{\ell\ell'} f_{i\ell'}^\dagger f_{i\ell}
    \right].
    \label{eq:HJ_MF_O_form}
\end{align}

We can gain further insight into the physical content of the mean-field theory by rewriting (see Appendix \ref{appendix:mft}) the $O_i^{\ell\ell'}$ mean fields in terms of the average filling per site, $n_i$, and the ``magnetization'' vector, $\vec{m}_i$, defined in Eq.~\eqref{eq:mia}. This leads to a particularly simple form at the SU($N$)-symmetric point $J_{\ell\ell'} = J$:
\begin{align}
    \tilde{\HH}_{J,\mathrm{MF}} &=
    N_c e_0 -8J \sum_i \vec{m}_i \cdot \vec{S}_i
    \notag \\
    &+ 4J\frac{(N-1)}{N} \sum_i \left( n_{i} - \frac{N}{2} \right) \hat{n}_{i}.
    \label{eq:HJ_MF_sun}
\end{align}
Here, $\vec{S}_i$ are the SU($N$) generators in Eq.~\eqref{eq:SUN generators} and $N_c e_0$ is a constant that depends on $\left\{ n_i, \vec{m}_i \right\}$. Hence, it is clear that a nonzero $\vec{m}_i$ signals the onset of SU($N$) symmetry-breaking order, as noted in Sec.~\ref{subsec:symm}.
In the usual $N=2$ case, a collinear spin density wave with nonzero $\vec{m}_i = e^{\ii \vec{Q}\cdot \vec{r}_i} \vec{m}$ lowers the symmetry of $\tilde{\HH}$ from the emergent group $G = \mathrm{SU(2)}$ to the subgroup $H = \mathrm{U(1)}$ formed by spin rotations around the ordering axis $\vec{m}$.
However, the situation becomes richer for $N>2$ because a FDW $\vec{m}_i = e^{\ii \vec{Q}\cdot \vec{r}_i} \vec{m}$ can have different residual symmetries $H$, and consequently realize different types of order, depending on the \emph{direction} of $\vec{m}$.

To illustrate this last point, consider the case $N=3$, for which the eight generators $S_i^a = \frac{1}{2} \sum_{\ell,\ell'} f_{i\ell}^\dagger \lambda^a_{\ell\ell'} f_{i\ell'}$ are specified by the Gell-Mann matrices $\lambda^a$.
Given an order parameter $\vec{m} \ne \vec{0}$, the symmetry $H$ 
of Eq.~\eqref{eq:HJ_MF_sun} is determined by finding the set of vectors $\vec{t}$ for which $\comm{ \vec{t} \cdot \vec{S}_i }{ \vec{m} \cdot \vec{S}_i } = 0$. This is equivalent to obtaining the centralizer of the matrix $M = \vec{m} \cdot \bm{\lambda} = \sum_a m_a \lambda^a$ in SU(3), i.e., the set of matrices $T = \vec{t} \cdot \bm{\lambda}$ that commute with $M$.
We can now contrast two configurations of $\vec{m}$ that couple to the different diagonal Gell-Mann matrices: (i) $M = m \lambda^3$ and (ii) $M = m \lambda^8$, where
\begin{align}
    \lambda^3 &= 
    \begin{pmatrix}
        1 & 0 & 0
        \\
        0 & -1 & 0
        \\
        0 & 0 & 0
    \end{pmatrix},
    &
    \lambda^8 &= \frac{1}{\sqrt{3}}
    \begin{pmatrix}
        1 & 0 & 0
        \\
        0 & 1 & 0
        \\
        0 & 0 & -2
    \end{pmatrix}.
    \label{eq:l3l8}
\end{align}
In case (i), the most general matrix that commutes with $M$ has the form $T = t_3 \lambda^3 + t_8 \lambda^8$. This implies that Eq.~\eqref{eq:HJ_MF_sun} is only invariant under global SU(3) rotations generated by $S = t_3 S^3 + t_8 S^8$, which can be decomposed into the product of transformations generated by $S^3$ and $S^8$ separately. Therefore, the residual symmetry is $H_1 = \mathrm{U(1)} \times \mathrm{U(1)}$.
By contrast, in case (ii), $M$ commutes with any matrix $T = \sum_{a=1}^3 t_a \lambda^a + t_8 \lambda^8$, so that Eq.~\eqref{eq:HJ_MF_sun} is invariant under SU(3) rotations generated by $S = \sum_{a=1}^{3} t_a S^a$, $S^8$, or any linear combination thereof. From this we conclude that the system now has a larger invariant subgroup $H_2 = \mathrm{SU(2)} \times \mathrm{U(1)}$, which leads to a different phase than (i).

\begin{figure}[t!]
    \centering
    \includegraphics[width=0.7\linewidth]{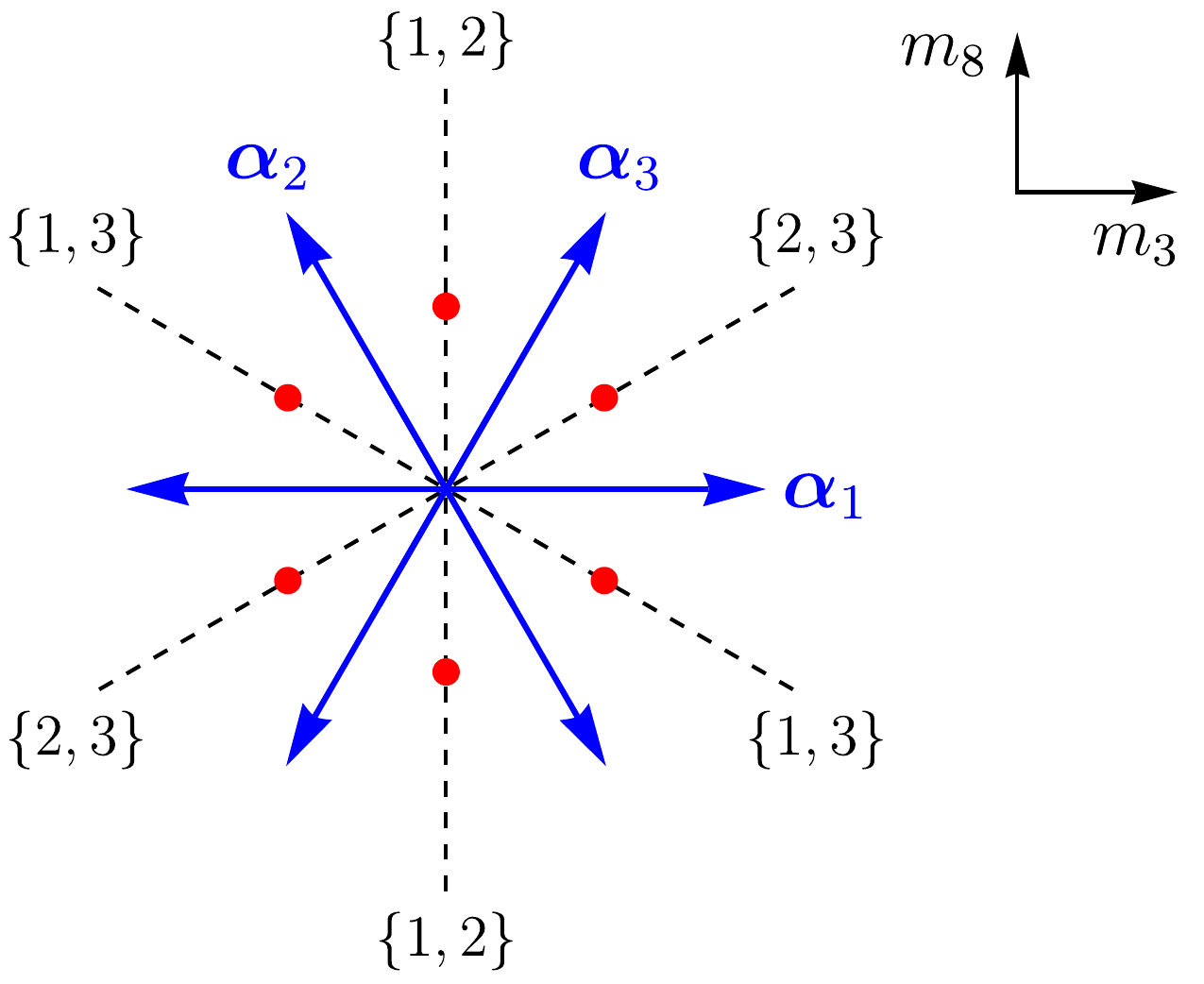}
    \caption{Schematic illustrating a set of singular order-parameter configurations $\vec{m} = (m_3,m_8)$ (red dots) that lead to a residual $H = \mathrm{SU(2)} \times \mathrm{U(1)}$ symmetry for the $N=3$ mean-field Hamiltonian in Eq.~\eqref{eq:HJ_MF_sun}. Every one of these configurations lies on top of a line that is perpendicular to a root of SU(3) (blue vectors), and each such line is labeled by the pair of flavors $\{ \ell, \ell' \}$ that is rotated under the $\mathrm{SU(2)}$ transformations in $H$.}
    \label{fig:m3m8}
\end{figure}

By generalizing this analysis (see Appendix~\ref{appendix:su3 proof}), we find that most mean-field configurations $\vec{m}$ yield the same residual symmetry as case (i). The exceptions all correspond to case (ii) and are identified as follows. Given a mean field $\vec{m}$, one can diagonalize the matrix $M = \vec{m} \cdot \bm{\lambda}$ via a unitary transformation $U$ to obtain $M' = U^\dagger M U = m_3' \lambda^3 + m_8' \lambda^8$. As depicted in Fig.~\ref{fig:m3m8}, the invariant subgroup will be $H_2$ if and only if $\vec{m}' = (m_3', m_8')$ is perpendicular to one of the three vectors $\bm{\alpha}_1 = (1,0)$, $\bm{\alpha}_2 = (-1, \sqrt{3})/2$, or $\bm{\alpha}_3 = (1, \sqrt{3})/2$ corresponding to the (positive) roots of SU(3) \cite{zee_book}. As shown in Appendix~\ref{appendix:su3 proof}, this condition is equivalent to $M$ having two degenerate eigenvalues.

%%%%%%%%%%%%%%%%%%%%%%%%%%%%%%%%%%%%%%%%%%%%%%%%%%%%%%%%%%%
%%%%%%%%%%%%%%%%%%%%%%%%%%%%%%%%%%%%%%%%%%%%%%%%%%%%%%%%%%%
%%%%%%%%%%%%%%%%%%%%%%%%%%%%%%%%%%%%%%%%%%%%%%%%%%%%%%%%%%%

\section{Mean-field results for $N=3$}
\label{sec:mft}

We now focus on the solution of the half-filled $N=3$ mean-field theory, which is the simplest case to admit symmetry-broken states with different residual symmetry groups $H$. 
As mentioned previously, we will treat $\tilde{\HH}_K$ in a fixed gauge sector $\{ u \}$ that implements a $\pi$-flux hopping background. With this choice, the system's unit cell is doubled from the outset, so that a staggered FDW order is no longer described by an ordering wavevector $\vec{Q} = (\pi,\pi)$, as in a zero-flux square lattice, but instead by $\vec{Q} = \bm{0}$ and sublattice-dependent mean fields. Assuming that this type of FDW is the leading instability out of the Dirac semimetal realized at low $J/K$, we restrict ourselves to states where $n_{i} = n_{\mu}$ and $\vec{m}_i = \vec{m}_{\mu}$ for sites $i$ in sublattice $\mu \in \left\{ A,B \right\}$.
We further set $\vec{m} = \vec{m}_A = -\vec{m}_B$, which follows from the invariance of the Hamiltonian under a spatial inversion that exchanges sublattices $A \leftrightarrow B$ and a global $\pi$ rotation $U = \exp(-i\pi S^a)$.
Altogether, this leaves us with nine independent mean-field parameters, which can be chosen to be the eight components of $\vec{m}$ plus $\Delta n_A = n_A - 3/2$, the deviation of the $A$-sublattice occupation from half filling.

The remainder of this section is structured as follows. In Sec.~\ref{subsec:mftn3_eqs}, we use the above assumptions to derive a specialized version of the theory laid out in Sec.~\ref{sec:mft_general}. Then, in Sec.~\ref{subsec:mftn3_phases+pd}, we examine the solutions of the mean-field equations for a specific, but physically motivated, type of anisotropy of the couplings $J_{\ell\ell'}$. After describing the zero-temperature phases in detail, we discuss the resulting phase diagram and potential effects beyond the scope of our mean-field theory.

%%%%%%%%%%%%%%%%%%%%%%%%%%%%%%%%%%%%%%%%%%%%%%%%%%%%%%%%%%%
%%%%%%%%%%%%%%%%%%%%%%%%%%%%%%%%%%%%%%%%%%%%%%%%%%%%%%%%%%%

\subsection{Mean-field equations}
\label{subsec:mftn3_eqs}

By imposing the assumptions outlined at the beginning of this section and performing a Fourier transform for a system with $N_c$ unit cells, we arrive at the mean-field Hamiltonian
\begin{equation}
    \tilde{\HH}_{\rm MF}
    =
    N_c e_0
    +\sum_{\vec k}
    \Psi_{\vec k}^\dagger
    \begin{pmatrix}
    \mathcal{J} & g_{\vec{k}} \, \mathds{1}_{3} \\
    g_{\vec{k}}^* \, \mathds{1}_{3} & -\mathcal{J}
    \end{pmatrix}
    \Psi_{\vec k},
    \label{eq:HMF_Neq3}
\end{equation}
where $g_\vec{k} = 4K(\sin k_x + \ii \cos k_y)$ is the hopping form factor in the $\pi$-flux sector, $\mathds{1}_{3}$ is the $3\times 3$ identity matrix, and
\begin{equation}
    \Psi_\vec{k}^\dagger = 
    \begin{pmatrix} 
    f_{\vec{k}A1}^\dagger & f_{\vec{k}A2}^\dagger & f_{\vec{k}A3}^\dagger & f_{\vec{k}B1}^\dagger & f_{\vec{k}B2}^\dagger & f_{\vec{k}B3}^\dagger \end{pmatrix}
\end{equation}
is a vector of operators $f_{\vec{k}\mu\ell}^\dagger$ that create fermions of flavor $\ell$ spread over the sites of sublattice $\mu$ with a periodicity given by the wavevector $\vec{k}$. Furthermore, we have
\begin{align}
    e_0 &= e_0' - 8\sum_{\ell, \ell' > \ell} J_{\ell\ell'} \left[ 
    \left( \frac{\Delta n_A}{3} \right)^2 + (M_{\ell\ell} + M_{\ell' \ell'}) \frac{\Delta n_A}{3}
    \right.
    \notag \\ &
    \phantom{=e_0' - 8\sum_{\ell}\sum_{\ell' > \ell} J_{\ell\ell'}}
    \left. 
    + M_{\ell\ell}M_{\ell' \ell'} - \abs{M_{\ell\ell'}}^2 
    \right],
    \label{eq:n3_e0}
    \\
    \mathcal{J}_{\ell'\ell} &=
    - \left( 1-\delta_{\ell'\ell} \right) 4J_{\ell\ell'} M_{\ell'\ell}
    \notag \\ &\phantom{=}
    + \delta_{\ell'\ell} \sum_{\ell\prpr \ne \ell} 4J_{\ell\ell\prpr} \left( \frac{\Delta n_A}{3} + M_{\ell\prpr\ell\prpr} \right),
    \label{eq:n3_Jmat}
\end{align}
with $e_0'$ being an irrelevant constant and $M = \vec{m} \cdot \bm{\lambda}$. The presence of the cross-terms in Eq.~\eqref{eq:n3_e0} hints at a possible intertwinement of FDW and CDW orders, which is indeed observed in the phase diagram of the model (see Sec.~\ref{subsec:mftn3_phases+pd}).

The solutions $\{ \Delta n_A, \vec{m} \}$ of the mean-field theory are obtained by fulfilling the following set of self-consistency equations:
\begin{align}
    \Delta n_A &= \frac{1}{N_c} \sum_{\vec{k}\ell} \expval{ f_{\vec{k}A\ell}^\dagger f_{\vec{k}A\ell} } - \frac{3}{2},
    \label{eq:DnA_selfconsistency}
    \\
    m_a &= \frac{1}{2N_c} \sum_{\vec{k}} \sum_{\ell\ell'} \expval{ f_{\vec{k}A\ell}^\dagger \lambda_{\ell\ell'}^a f_{\vec{k}A\ell'} },
    \label{eq:ma_selfconsistency}
\end{align}
with $a=1,\ldots, 8$. Whereas different types of FDWs can be diagnosed by the vector $\vec{m}$, $\Delta n_A$ serves as an order parameter for a CDW, since the occupation of $A$ and $B$ sites differs whenever $\Delta n_A \ne 0$.

We can translate a solution $\{ \Delta n_A, \vec{m} \}$ of the mean-field equations into flavor-dependent densities $n_{\mu\ell}$ by means of the transformation (see Appendix~\ref{appendix:mft})
\begin{align}
    \begin{pmatrix}
        n_{\mu1} \\ n_{\mu2} \\ n_{\mu3}
    \end{pmatrix}
    =
    \frac{1}{2} \mathds{1} \pm
    \begin{pmatrix}
        1/3 & 1 & 1/\sqrt{3}
        \\
        1/3 & -1 & 1/\sqrt{3}
        \\
        1/3 & 0 & -2/\sqrt{3}
    \end{pmatrix}
    \begin{pmatrix}
        \Delta n_A \\ m_3 \\ m_8
    \end{pmatrix},
    \label{eq:nil N=3}
\end{align}
where the upper and lower signs refer to sublattices $\mu=A$ and $B$, respectively.
In particular, this transformation shows that, if $\vec{m}$ lies along one of the dashed lines $\{ \ell, \ell' \}$ in Fig.~\ref{fig:m3m8}, then two of the flavor-resolved densities are identical: $n_{\mu\ell} = n_{\mu\ell'}$. This is a manifestation of the enhanced SU(2)$\times$U(1) symmetry obtained along those special directions in $\vec{m}$-space.

%%%%%%%%%%%%%%%%%%%%%%%%%%%%%%%%%%%%%%%%%%%%%%%%%%%%%%%%%%%
%%%%%%%%%%%%%%%%%%%%%%%%%%%%%%%%%%%%%%%%%%%%%%%%%%%%%%%%%%%

\subsection{Mean-field phases and phase diagram in the presence of a partial layer anisotropy}
\label{subsec:mftn3_phases+pd}

To proceed, we consider a specific anisotropy of the interlayer couplings: $J_{12}=J_{23}= J$ and $J_{13}= J'$. This choice is justified if the system has a mirror symmetry with respect to the middle layer, $\ell = 2$ [see Fig.~\ref{fig:somodel}(b)]. The Hamiltonian $\tilde{\HH}$ then features an emergent $G = \mathrm{SU(2)}_{1,3} \times \mathrm{U(1)}$ symmetry, where $\mathrm{SU(2)}_{1,3}$ refers to the group of global SU(2) transformations that act on the equivalent flavors $\ell = 1,3$ and are generated by $\{ \tilde{S}^3, S^4, S^5 \}$ with $\tilde{S}_3 = (S^3 + \sqrt{3} S^8)/2$. The remaining U(1) subgroup in $G$ is generated by
\begin{equation}
    \tilde{S}^8 = \frac{1}{2}(\sqrt{3} S^3 - S^8) = \frac{1}{2} \sum_{i\ell} f_{i\ell}^\dagger \tilde{\lambda}_8^{\ell\ell'} f_{i\ell'},
\end{equation}
with $\tilde{\lambda}_8 = \diag (1,-2,1)/\sqrt{3}$ being an analog of $\lambda_8$ that treats flavors $\ell=1,3$ on equal footing. At the isotropic point $J'/J = 1$, the symmetry group of the Hamiltonian is enhanced to $G = \mathrm{SU(3)}$.

We solved Eqs.~\eqref{eq:DnA_selfconsistency} and \eqref{eq:ma_selfconsistency} numerically at temperature $T=0$ for a wide range of parameters $\{K, J, J'\}$ on square clusters with $N_c = L^2$ unit cells and linear system sizes $L \in \{ 80, 160, 320, 640, 1280 \}$. As a result, we found that, although a generic mean-field solution $\{ \Delta n_A, \vec{m} \}$ can feature more than two nonzero components $m_a$, there is \emph{always} an element in $G$ that connects it to a state $\{ \Delta n_A, \vec{m}' \}$ with $m'_a = 0$ for $a\ne3,8$\footnote{For $J'/J\ne 1$, this implies that the matrix $M=\vec{m}\cdot \bm{\lambda}$ can be written as a linear combination of just $\lambda_a$ with $a=3,4,5,8$. For $J'/J=1$, $M$ can receive nonvanishing contributions from all Gell-Mann matrices.}.
Thanks to this property, we can gain analytical insight into the nature of the fermionic spectrum of the different mean-field phases. Indeed, if we set $M = m_3 \lambda_3 + m_8 \lambda_8$, Eq.~\eqref{eq:HMF_Neq3} decomposes into $2\times2$ blocks of fixed momentum $\vec{k}$ and flavor $\ell$:
\begin{equation}
    \tilde{\HH}_{\rm MF}
    =
    N_c e_0
    +\sum_{\vec k}\sum_{\ell=1}^3
    \Psi_{\vec k\ell}^\dagger
    \begin{pmatrix}
    \Delta_{\ell} & g_{\vec k} \\
    g_{\vec k}^\ast & -\Delta_{\ell}
    \end{pmatrix}
    \Psi_{\vec k\ell},
    \label{eq:HMF_blocks_diagNeq3}
\end{equation}
where $\Psi_{\vec k\ell}^\dagger = \begin{pmatrix} f^\dagger_{\vec kA\ell} & f^\dagger_{\vec kB\ell} \end{pmatrix}$ and
\begin{align}
    \Delta_{\ell} &=
    4 \sum_{\ell'\neq \ell} J_{\ell\ell'} \Bigl(
    \frac{\Delta n_A}{3} + M_{\ell'\ell'}
    \Bigr).
    \label{eq:Deltaell_generalJ}
\end{align}
The straightforward diagonalization of Eq.~\eqref{eq:HMF_blocks_diagNeq3} then yields six bands with dispersion
\begin{equation}
    \varepsilon_{\vec k\ell \pm} = \pm \sqrt{|g_{\vec k}|^2+\Delta_\ell^2}.
    \label{eq:MF_spectrum}
\end{equation}
Hence, it is clear that a nonzero $\Delta_\ell$ gaps a pair of Dirac cones present in the noninteracting spectrum of Eq.~\eqref{eq:HK_spectrum}. Note that the structure of the Gell-Mann matrices $\lambda_3$ and $\lambda_8$ gives rise to a nontrivial interplay between $\Delta n_A$ and $\vec{m}$ in Eq.~\eqref{eq:Deltaell_generalJ}, such that increasing $\abs{\Delta n_A}$ or $\abs{\vec{m}}$ does not necessarily lead to an increase in $\abs{\Delta_\ell}$.

%%%%%%%%%%%%%%%%%%%%%%%%%%%%%%%%%%%%%%%%%%%%%%%%%%%%%%%%%%%

\subsubsection{Phases at $T=0$}
\label{subsubsec:phases}

After carefully extrapolating the finite-size solutions of Eqs.~\eqref{eq:DnA_selfconsistency} and \eqref{eq:ma_selfconsistency} to the thermodynamic limit $L\to \infty$, we found that the system realizes \emph{five} different phases at temperature $T=0$.

The first phase is a Dirac semimetal, which is characterized by $(\Delta n_A, \vec{m}) = (0, \bm{0})$ and has a total of six Dirac cones (two per flavor $\ell$). This is a maximally symmetric phase, since it preserves both the $\mathbb{Z}_2$ sublattice symmetry and the symmetry $G$ in flavor space.

\begin{figure}[t!]
    \centering
    \includegraphics[width=\linewidth]{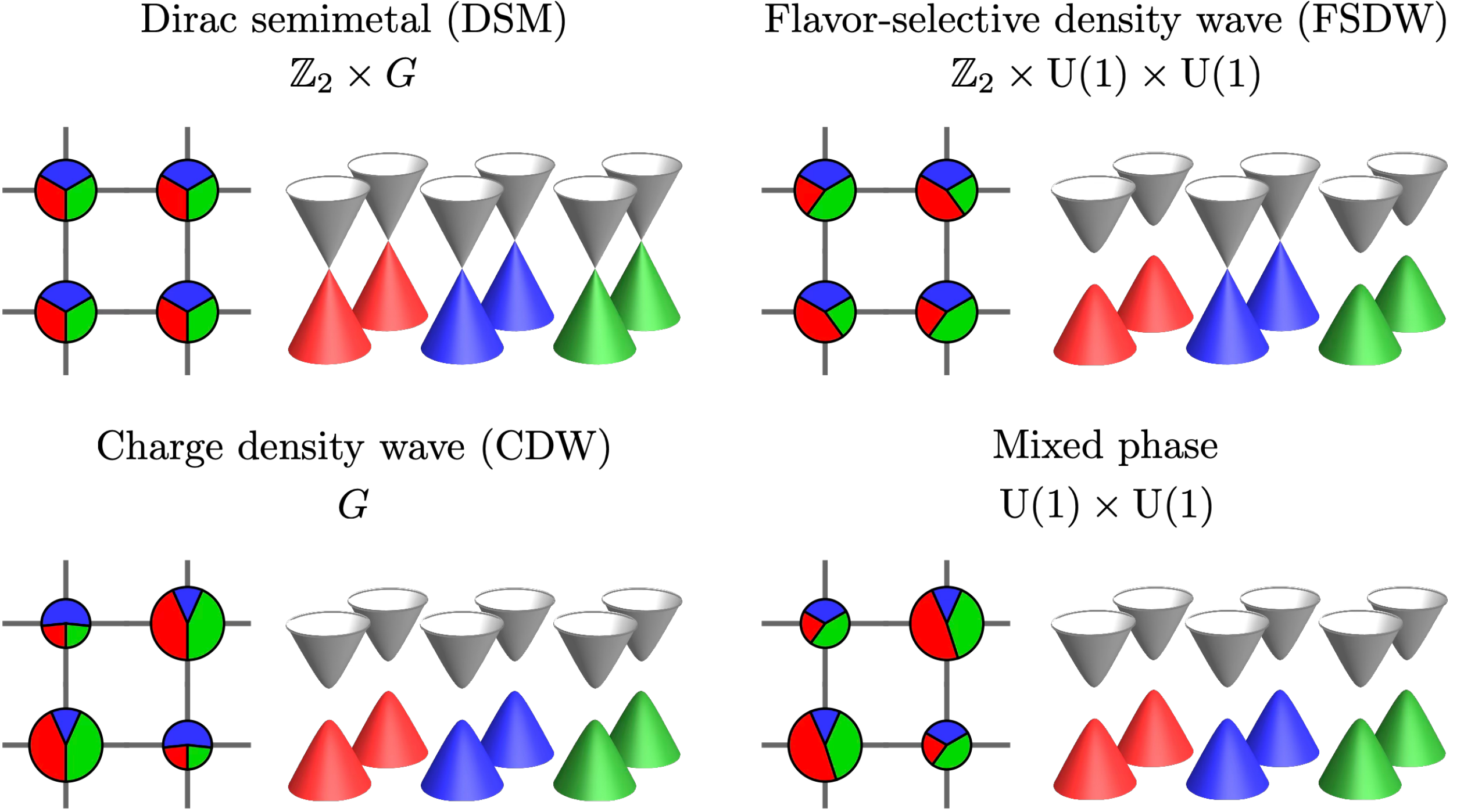}
    \caption{Four different phases obtained as mean-field solutions of the $N=3$ Hubbard model with anisotropic interactions $J_{12}=J_{23}=J$ and $J_{13}=J'$ at temperature $T=0$. Each phase is identified by its residual symmetry and two schematics.
    On the left, we depict the spatial distribution of the flavor-resolved densities $n_{\mu\ell}$ in terms of pie charts where the fractions of $\ell=1,2,3$ fermions are represented in red, blue and green, respectively. The diameter of any given disk is proportional to the total charge $n_\mu = \sum_{\ell} n_{\mu\ell}$ at the corresponding lattice site.
    The right schematics illustrate the fermionic spectrum of each phase at low energy, focusing on whether the Dirac cones present in the noninteracting limit are gapped or not.}
    \label{fig:phases}
\end{figure}

The second phase corresponds to solutions that develop CDW order while preserving $G$. Note that the first condition implies $\Delta n_A \ne 0$, but the second only requires $\vec{m}=\bm{0}$ at the SU(3)-symmetric point $J'/J = 1$. Away from this limit, the $\SUtwocrossUone$ symmetry of the Hamiltonian allows for a nonzero $\vec{m}$ along the $\{1,3\}$ dashed line in Fig.~\ref{fig:m3m8}. Substituting $m_a = m (\sqrt{3} \delta_{a3} - \delta_{a8})/2$ into Eq.~\eqref{eq:nil N=3}, we find
\begin{align}
    n_{\mu1} = n_{\mu3} &= \frac{1}{2} \pm \frac{\Delta n_A}{3} \pm \frac{1}{\sqrt{3}} m,
    \notag \\
    n_{\mu2} &= \frac{1}{2} \pm \frac{\Delta n_A}{3} \mp \frac{2}{\sqrt{3}} m,
    \label{eq:nil CDW}
\end{align}
where, as before, the upper (lower) sign refers to sublattice $\mu=A$ ($\mu=B$). This shows that, in a CDW solution with $m\ne 0$, the spatial modulation of $n_{\mu2}$ is out of phase with that of $n_{\mu1} = n_{\mu3}$, so that one sublattice is predominantly occupied by $\ell = 2$ fermions while the other shows a higher density of $\ell=1$ and $3$ fermions. As evidenced by Eq.~\eqref{eq:Deltaell_generalJ} and illustrated in Fig.~\ref{fig:phases}, the fermionic spectrum of this phase is fully gapped.

The third phase is a FDW that spontaneously breaks the symmetry $G$ down to $H_1 = \UonecrossUone$ and lacks CDW order ($\Delta n_A = 0$). From our numerical analysis, we find that this type of solution arises when $\vec{m}$ is parallel to $\bm{\alpha}_3$ (see Fig.~\ref{fig:m3m8}) or lies along an equivalent direction under the symmetry $G$. To reveal further information about the structure of this phase in real space, we set $m_a = m(\delta_{a3} + \sqrt{3}\delta_{a8})/2$ in Eq.~\eqref{eq:nil N=3}. The result
\begin{align}
    \left( n_{\mu1}, n_{\mu2}, n_{\mu3} \right) =
    \left( \frac{1}{2} \pm m, \,\frac{1}{2}, \,\frac{1}{2} \mp m \right)
    \label{eq:nil FSDW}
\end{align}
indicates that only \emph{two} of the fermionic flavors ($\ell=1,3$) have nonuniform densities for $m\ne 0$. Since transformations in $G = \SUtwocrossUone$ do not affect $n_{\mu2}$, we conclude that this is a generic feature of the phase in question, which we therefore term a \emph{flavor-selective} density wave (FSDW).
From Eq.~\eqref{eq:Deltaell_generalJ}, we verify that four of the six original Dirac cones are gapped in the FSDW. The last pair of gapless Dirac cones corresponds to the $\ell=2$ fermions, which remain fully delocalized, in contrast to their $\ell=1,3$ counterparts. Hence, the FSDW is an example of a flavor-selective insulator \cite{delre18,tusi22}.

The fourth phase encompasses solutions in which a FDW with residual $H_1 = \mathrm{U(1)} \times \mathrm{U(1)}$ symmetry coexists with CDW order. As such, this ``mixed'' phase differs from the FSDW in that it lacks a $\mathbb{Z}_2$ symmetry given by the product of an elementary translation with a permutation of the equivalent flavors $\ell = 1,3$. In our numerical calculations, we find that the mixed phase emerges for generic magnetizations $\vec{m}$ which, when mapped onto $m_3$-$m_8$ plane via a transformation in $G$, are neither parallel nor perpendicular to the root vectors in Fig.~\ref{fig:m3m8}. As illustrated in Fig.~\ref{fig:phases}, the fermionic spectrum of this phase is fully gapped.

Finally, when $J'/J=1$, the enhanced SU(3) symmetry of the Hamiltonian enables the emergence of another ordered phase: a FDW with residual $H_2 = \mathrm{SU(2)}\times\mathrm{U(1)}$ symmetry. As discussed at the end of Sec.~\ref{sec:mft_general}, this type of state occurs when $\vec{m}$ is parallel to special directions in the $m_3$-$m_8$ plane (dashed lines in Fig.~\ref{fig:m3m8}) or can be mapped onto them via a unitary transformation. The explicit solutions of the mean-field equations show that this fifth phase also displays a nonvanishing CDW component $\Delta n_A \ne 0$. Thus, the $\mathrm{SU(2)}\times\mathrm{U(1)}$ FDW in question resembles the CDW phase described above, both in the fully gapped nature of its spectrum \cite{miyatake10} and the spatial dependence of the densities $n_{\mu\ell}$ (up to rotations in flavor space). The reason for this similarity will be evinced in the analysis of the phase diagram below.

%%%%%%%%%%%%%%%%%%%%%%%%%%%%%%%%%%%%%%%%%%%%%%%%%%%%%%%%%%%

\begin{figure*}[t!]
    \centering
    \includegraphics[width=\linewidth]{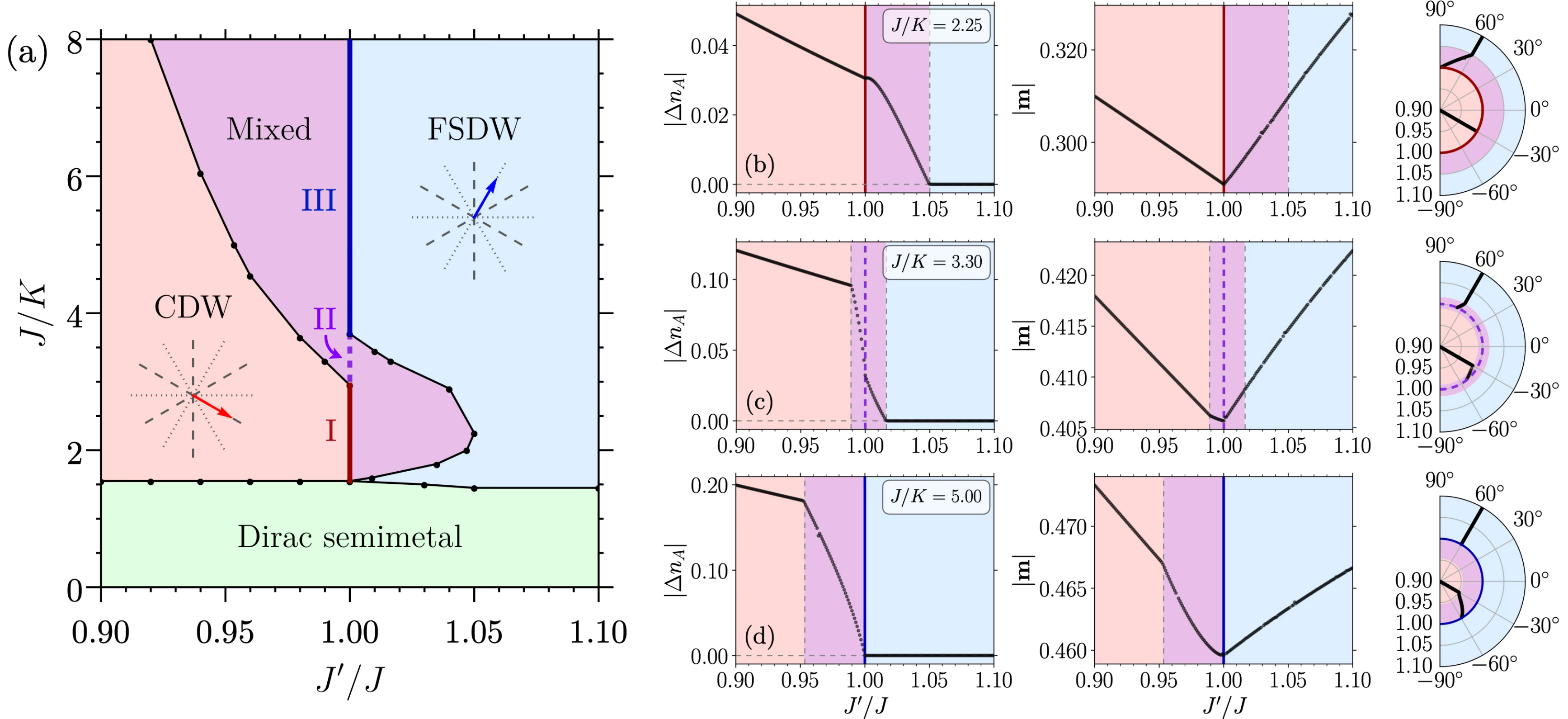}
    \caption{Mean-field results for the $\pi$-flux, $N=3$ Hubbard model at half filling and with anisotropic interlayer (Hubbard) interactions $J_{12}=J_{23}=J$ and $J_{13}=J'$.
    (a) Phase diagram obtained after finite-size scaling. The insets in the CDW and FSDW phases depict the direction within the $m_3$-$m_8$ plane (see also Fig.~\ref{fig:m3m8}) is selected by the anisotropy $J'/J$. On the SU(3)-symmetric line $J'/J=1$, the thick red line I marks a phase where CDW order coexists with a FDW with a residual SU(2)$\times$U(1) symmetry. Meanwhile, regions II and III highlight extensions of the mixed and FSDW phases, respectively.
    (b-d) Evolution of mean-field parameters along horizontal cuts of the phase diagram corresponding to $J/K = 2.25$, $3.30$ and $5.00$. The colors in the background of these plots refer to the phases in (a). The left and middle panels display the absolute value of the CDW order parameter, $\Delta n_A = n_A -3/2$, and the norm of the magnetization vector, $\vec{m}$, respectively. The right panels depict how the direction of $\vec{m}$ within the $m_3$-$m_8$ plane with $J'/J$ (radial component) for a mean-field solution with $m_a=0$ for $a\ne 3,8$. At the isotropic point $J'/J=1$, solutions related by a 60° rotation in the $m_3$-$m_8$ plane are equivalent under the SU(3) symmetry of the Hamiltonian.}
    \label{fig:mfpd}
\end{figure*}

\subsubsection{Phase diagram}

The $T=0$ phase diagram of the $N=3$ mean-field theory is shown in Fig.~\ref{fig:mfpd}(a). Starting in the noninteracting limit $J/K = 0$ of the SU(3)-symmetric line $J'/J = 1$ and increasing the interaction strength, we observe that the Dirac semimetal is destabilized at $(J/K)_{c1} \approx 1.55$. The ensuing phase, which is represented by a dark red line (I) in the phase diagram, is the $\mathrm{SU(2)}\times\mathrm{U(1)}$ FDW described at the end of Sec.~\ref{subsubsec:phases}.

The influence of a $J'/J$ anisotropy on the $\mathrm{SU(2)}\times\mathrm{U(1)}$ FDW is illustrated in Fig.~\ref{fig:mfpd}(b), which contains mean-field results at fixed $J/K = 2.25$.
On the one hand, when $J'/J < 1$, the anisotropy lifts the degeneracy of $\mathrm{SU(2)}\times\mathrm{U(1)}$ FDW states with different $\vec{m}$ and drives the system into the CDW phase. As evidenced by Eq.~\eqref{eq:nil CDW}, this occurs because CDW states minimize the interaction between flavors $\{ 1,2 \}$ and $\{ 2,3 \}$, which are subject to the stronger repulsion $J>J'$.
On the other hand, when $J'/J > 1$, the interactions promote the minimization of contact between $\ell = 1,3$ fermions. For sufficiently large $J'/J$, this gives rise to the FSDW phase with the density profile of Eq.~\eqref{eq:nil FSDW}, in qualitative agreement with previous dynamical mean-field theory (DMFT) calculations for the zero-flux model \cite{miyatake10}.
However, rather than being abrupt, the transition to the FSDW occurs through an intermediate mixed phase in which $\Delta n_A$ is gradually suppressed as $\vec{m}$ distances itself from the directions with the higher $\mathrm{SU(2)}\times\mathrm{U(1)}$ symmetry.
The existence of an intervening mixed phase can be attributed to the fact that, unlike the SU(2)$\times$U(1) FDW, the FSDW does not have a fully gapped spectrum (see Fig.~\ref{fig:phases}). Thus, it only becomes energetically competitive once $J'/J$ is large enough for the closing of the gap $\abs{\Delta_2}$ in Eq.~\eqref{eq:Deltaell_generalJ} to be compensated by an enhancement of $\abs{\Delta_{1}}$ and $\abs{\Delta_{3}}$.
As one approaches the boundary with the Dirac semimetal, this compensation requires smaller and smaller anisotropies $J'/J$ because the gaps of SU(2)$\times$U(1) FDW scale with $[J/K-(J/K)_{c1}]$.

Returning to the SU(3)-symmetric line, we find that further increase in $J/K$ leads to two additional phase transitions at $(J/K)_{c2} \approx 2.95$ and $(J/K)_{c3} \approx 3.70$.
These transitions give rise to extensions of the mixed and FSDW phases, marked II and III in Fig.~\ref{fig:mfpd}(a), in which residual $\UonecrossUone$ and $\mathbb{Z}_2 \times \UonecrossUone$ symmetries are now obtained by spontaneously breaking the SU(3) symmetry.
As above, the presence of the mixed phase between regions I and III can be rationalized in terms of the qualitative change from a fully to partially gapped fermionic spectrum.

The responses of phases II and III to an interlayer anisotropy, shown in Figs.~\ref{fig:mfpd}(c,d) for $J/K = 3.30$ and $5.00$, can be understood in the same terms as Fig.~\ref{fig:mfpd}(b).
For $J'/J<1$, the anisotropy favors FDWs with higher spatial overlap between flavors $\ell=1,3$. This causes $\vec{m}$ to gradually move toward the symmetric $\{ 1,3 \}$ line in the $m_3$-$m_8$ plane and eventually stabilizes the CDW phase. Conversely, when $J'/J>1$, the mixed phase evolves continuously into a FSDW because the latter minimizes the interaction between $\ell=1,3$ fermions.

The fact that the polar plots of Figs.~\ref{fig:mfpd}(b-d) show a discontinuity in the orientation of $\vec{m}$ when $J'/J$ moves through the SU(3)-symmetric point indicates the system undergoes a first-order phase transition\footnote{This is analogous to the first-order phase transition observed, e.g., in a two-dimensional Ising ferromagnet when the magnitude of a longitudinal external field, $h$, is scanned through $h=0$ below the critical temperature $T_c$. At $h=0$, the two ordered ground states of the model are equivalent in the same way as the ends of the lines converging to $J'/J=1$ in the polar plots of Figs.~\ref{fig:mfpd}(b-d).} on crossing any of the phases I, II, or III in Fig.~\ref{fig:mfpd}(a).
All other phase transitions are found to be continuous within our mean-field theory. Notably, this includes the cascade of transitions realized at fixed $J'/J = 1$, where the least symmetric phase is II. In particular, the sequence $\mathrm{DSM} \!\to\! \mathrm{I} \!\to\! \mathrm{II}$ corresponds to a two-step symmetry-breaking transition \cite{torres20,niemi21} in which the residual symmetry groups evolve according to $\mathrm{SU(3)} \to \mathrm{SU(2)}\times\mathrm{U(1)} \to \UonecrossUone$. By contrast, the $\mathrm{III} \to \mathrm{II}$ transition only involves the breaking of a $\mathbb{Z}_2$ symmetry.

%%%%%%%%%%%%%%%%%%%%%%%%%%%%%%%%%%%%%%%%%%%%%%%%%%%%%%%%%%%

\subsubsection{Beyond mean-field theory}

To gauge the effect of \emph{local} fluctuations that are beyond the scope of our mean-field theory, we can compare our results to a previous DMFT study of the zero-flux version of the $N=3$ Hubbard model at hand \cite{miyatake10}. This study similarly found that a large $J'/J$ anisotropy drives\footnote{In the $0$-flux case, this occurs for any $J/K>0$, because the metallic phase realized at $J/K=0$ has a nesting instability at infinitesimal $J/K$ \cite{honerkamp04}.} the system into a ``color density wave'' ($J'/J<1$) or ``color-selective antiferromagnet'' ($J'/J>1$), corresponding precisely to the CDW and FSDW introduced here.
However, when analyzing data obtained by scanning $J/J' \in [0,2]$ at fixed $J'/K = 5$, Miyatake \textit{et al.} \cite{miyatake10} reported a first-order phase transition between the CDW and FSDW phases at $J'/J \approx 1$, with no reference to an intermediate mixed phase. Taken at face value, this suggests that the temporal fluctuations captured by DMFT may disfavor or even melt the mixed phase. However, a more thorough analysis is clearly needed to validate this interpretation. Such an analysis could, e.g., entail more extensive DMFT simulations (also of the $\pi$-flux model) focused on the vicinity of the SU(3)-symmetric point.

By contrast, we only expect \emph{spatial} fluctuations to qualitatively change our results above zero temperature, where the Hohenberg-Mermin-Wagner theorem \cite{hohenberg67,mermin66} forbids the establishment of long-range order from a spontaneously broken continuous symmetry. This determines that the FSDW phase is destroyed for any $T>0$. Meanwhile, when approaching the mixed phase from elevated temperatures, the system should display a thermal phase transition associated with the breaking of a $\mathbb{Z}_2$ symmetry. True long-range order from the breaking of $G$ is only achieved for $T\to 0$.

%%%%%%%%%%%%%%%%%%%%%%%%%%%%%%%%%%%%%%%%%%%%%%%%%%%%%%%%%%%
%%%%%%%%%%%%%%%%%%%%%%%%%%%%%%%%%%%%%%%%%%%%%%%%%%%%%%%%%%%
%%%%%%%%%%%%%%%%%%%%%%%%%%%%%%%%%%%%%%%%%%%%%%%%%%%%%%%%%%%

\section{Mapping back to the original degrees of freedom: Magnetic fragmentation and nonlocal orders}
\label{sec:mapback}

The mean-field phases discussed in Sec.~\ref{sec:mft} are described in terms of the parameters $\{\Delta n_A,\vec{m}\}$. To identify their physical content in terms of the original degrees of freedom, one must map these quantities back to the spin-orbital variables of Eqs.~\eqref{eq:HsoK} and \eqref{eq:HsoJ}. The key ingredient is the local $\mathbb{Z}_2$ gauge redundancy in Eq.~\eqref{eq:fermion gauge}: local densities $f_{i\ell}^\dagger f_{i\ell}$ are gauge invariant, whereas nondiagonal bilinears $f_{i\ell}^\dagger f_{i\ell'}$ with $\ell\neq \ell'$ are not. As a result, the diagonal components of $M=\vec{m}\cdot\bm{\lambda}$ map onto local order parameters via $\sigma_{i\ell}^z = 1-2f_{i\ell}^\dagger f_{i\ell}$ \cite{vijayvargia23}. For the diagonal SU(3) generators, one finds
\begin{align}
    S_i^3 &= \frac{1}{2}\left( \hat{n}_{i1} - \hat{n}_{i2} \right)
    = \frac{1}{4}\left( \sigma_{i2}^z - \sigma_{i1}^z \right), \notag\\
    S_i^8 &= \frac{1}{2\sqrt{3}}\left( \hat{n}_{i1} + \hat{n}_{i2} - 2\hat{n}_{i3} \right)
    = \frac{1}{4\sqrt{3}}\left( 2\sigma_{i3}^z - \sigma_{i1}^z - \sigma_{i2}^z \right),
    \label{eq:map_local_diag}
\end{align}
while the total density satisfies
\begin{equation}
    \hat{n}_i - \frac{3}{2}
    = -\frac{1}{2}\left( \sigma_{i1}^z + \sigma_{i2}^z + \sigma_{i3}^z \right).
    \label{eq:map_cdw}
\end{equation}
Therefore, a nonzero $\Delta n_A$ corresponds to a staggered order in the total magnetization per column $Z_i = \sum_\ell \expval{\sigma_{i\ell}^z}$. Furthermore, after spontaneous symmetry breaking, $\vec{m}$ can point along any of the directions detailed in Sec. \ref{subsec:mftn3_phases+pd}; any diagonal component of $M$ gives rise to local order between the on-site magnetizations $\expval{ \sigma_{i\ell}^z }$ in a fixed column $i$.
Corresponding to all the phases discussed in Sec. \ref{subsubsec:phases}, we depict the ordering patterns of the spins $\sigma_{i\ell}^z$ on the three layers in Fig.~\ref{fig:localorder_phases}.
By contrast, nondiagonal components of $M$ can only be detected through gauge-invariant string correlators as we will describe in the following. 

%%%%%%%%%%%%%%%%%%%%%%%%%%%%%%%%%%%%%%%%%%%%%%%%%%%%%%%%%%%
%%%%%%%%%%%%%%%%%%%%%%%%%%%%%%%%%%%%%%%%%%%%%%%%%%%%%%%%%%%

\subsection{Nonlocal string order}

Suppose that the ground state $\ket{\psi_u}$ of $\tilde{\HH}$ within a fixed gauge sector $\left\{ u \right\}$ displays long-range order in a nondiagonal channel, i.e.,
\begin{equation}
    m_a^u = \frac{1}{N_s}\sum_i (-1)^i \expval{S_i^a}{\psi_u} \neq 0
    \label{eq:mua_def}
\end{equation}
for some $a \neq 3,8$. Since the corresponding $S_i^a$ operator is not invariant under Eq.~\eqref{eq:fermion gauge}, Elitzur's theorem \cite{elitzur75} implies that the physical ground state $\ket{\psi} = P \ket{\psi_u}$ obeys $\sum_i (-1)^i \expval{S_i^a}{\psi}=0$. However, as shown in Appendix~\ref{appendix:string}, the order captured by Eq.~\eqref{eq:mua_def} survives in the form of gauge-invariant string correlations,
\begin{equation}
	C_\gamma^a \left(\vec{r}_i, \vec{r}_j\right)
    =
    \expval{
    S_{i}^a \,
    \hat{B}_\gamma^a \left(\vec{r}_i, \vec{r}_j\right) \,
    S_{j}^a
    }{\psi},
	\label{eq:Ca def}
\end{equation}
where
\begin{equation}
	\hat{B}_\gamma^a \left(\vec{r}_i, \vec{r}_j\right)
    =
    \prod_{(mn)\in \gamma} \hat{u}_{\ell,mn}\hat{u}_{\ell',mn}.
	\label{eq:Bdef}
\end{equation}
Here, $\gamma$ denotes an arbitrary path between the positions $\vec{r}_i$ and $\vec{r}_j$, and we suppress the bond label $\alpha$ on $\hat{u}_{\ell,mn}^\alpha$ for conciseness. The pair of layer indices $\{\ell,\ell'\}$ is determined by the generator $S^a$: $\{1,2\}$ for $a=1,2$, $\{1,3\}$ for $a=4,5$, and $\{2,3\}$ for $a=6,7$. Thus, every nondiagonal SU(3) channel is associated with a specific type of string operator \eqref{eq:Bdef}. This is in contrast with the existence of a single string operator in the SU(2) case observed in bilayer model \cite{vijayvargia23}. For general SU($N$), there will be $N(N-1)/2$ different $\hat{B}_\gamma^a$ operators: one for each (unordered) pair $\{\mu,\nu\}$ of SU($N$) indices $\mu,\nu = 1,\ldots,N$.

%%%%%%%%%%%%%%%%%%%%%%%%%%%%%%%%%%%%%%%%%%%%%%%%%%%%%%%%%%%
%%%%%%%%%%%%%%%%%%%%%%%%%%%%%%%%%%%%%%%%%%%%%%%%%%%%%%%%%%%

\subsection{Phases for $J_{13}=J' \neq J$}

For the partially anisotropic problem investigated in Sec.~\ref{subsec:mftn3_phases+pd}, the fixed-flux Hamiltonian $\tilde{\HH}$ has the reduced symmetry $G=\SUtwocrossUone$. Since only flavors $\ell=1,3$ are related by continuous SU(2) transformations, the only nondiagonal channels $a$ that can be generated by symmetry are those associated with $S^4$ and $S^5$. Accordingly, the anisotropic problem admits only one family of string order parameters, namely $C_\gamma^4$ and $C_\gamma^5$, built from the pair of layers $\{1,3\}$.

\begin{figure}[t!]
    \centering
    \includegraphics[width=\linewidth]{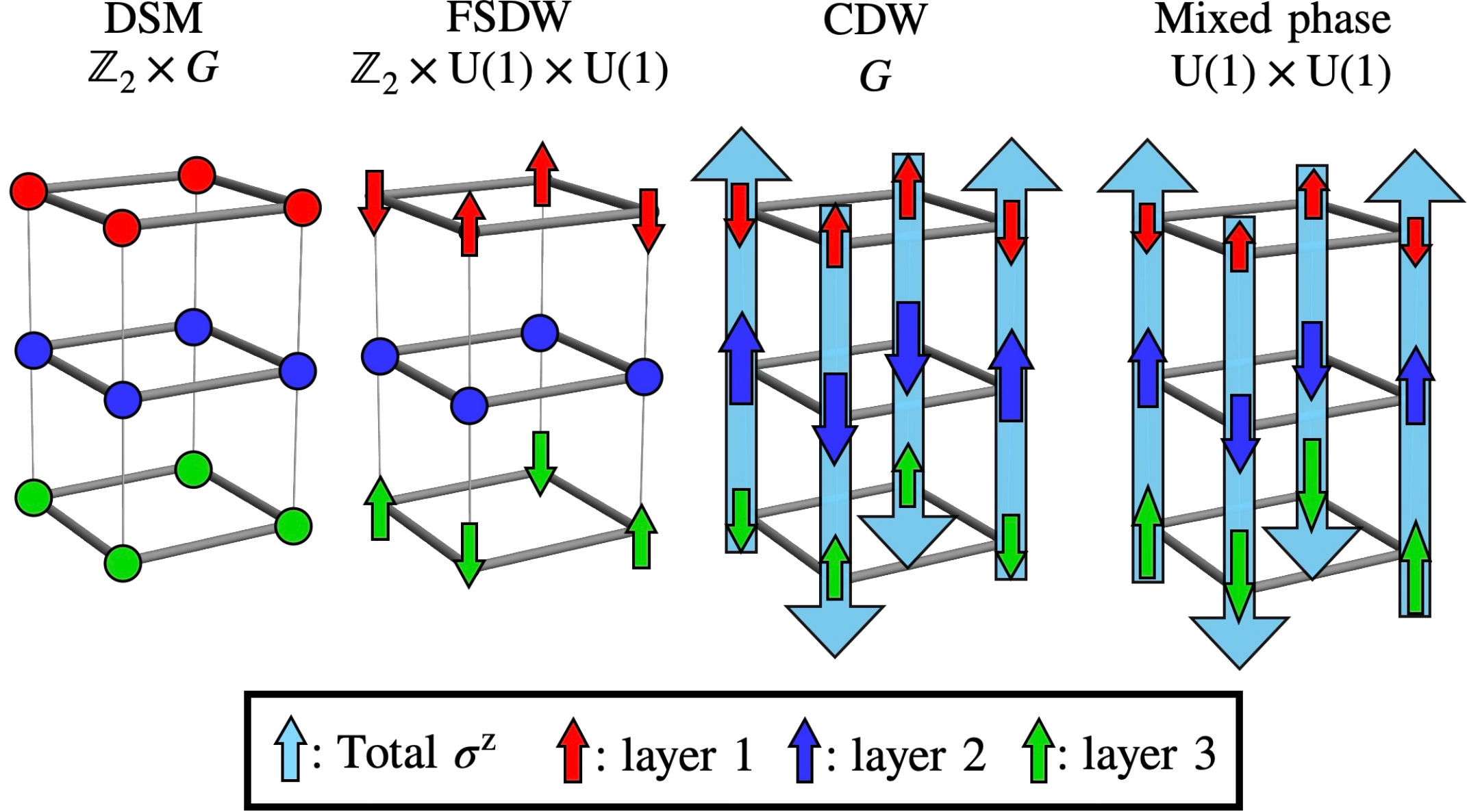}
    \caption{Representative ordering patterns for the four mean-field phases depicted in Fig.~\ref{fig:phases}. Each phase is illustrated on a $2\times2$ cluster of the three-layer lattice. Small arrows in red, dark blue and green denote the on-site magnetizations $\langle \sigma_{i\ell}^z \rangle = 1-2n_{i\ell}$ on layers $\ell=1,2,3$, respectively. In the DSM, all local moments vanish. The FSDW phase has opposite moments on layers 1 and 3, with a vanishing column magnetization $Z_i = \sum_{\ell}\langle \sigma_{i\ell}^z \rangle$. In the CDW phase, a symmetry-allowed staggered order within each column $i$ coexists with long-range antiferromagnetic order in $Z_i$, represented by the large light-blue arrows. The mixed phase contains both a relative layer polarization and a nonzero staggered total magnetization on the corresponding columns.}
    \label{fig:localorder_phases}
\end{figure}

As explained in Sec.~\ref{subsubsec:phases}, the CDW phase preserves the full symmetry $G$ while breaking the $\mathbb{Z}_2$ symmetry associated with the product of an elementary lattice translation times a $1\leftrightarrow3$ flavor exchange. Due to the property $\Delta n_A \ne 0$, this phase shows long-range antiferromagnetic order in the magnetizations per column $Z_i$. Furthermore, it displays a symmetry-allowed imbalance between the local magnetizations $\expval{\sigma_{i\ell}^z}$ in a fixed column $i$. Indeed, Eq.~\eqref{eq:nil CDW} implies $\expval{\sigma_{i1}^z}  = \expval{\sigma_{i3}^z} \ne \expval{\sigma_{i2}^z}$ whenever the symmetry-allowed diagonal component of $\vec{m}$ is nonzero.

The emergence of antiferromagnetic order in the $Z_i$ degrees of freedom despite the absence of an intralayer Ising coupling is most simply understood in the strongly anisotropic regime $J'\rightarrow0$ and $K/J\rightarrow 0$ [top-left corner of Fig. \ref{fig:phases}(a)]. In this limit, the dominant interlayer term $\HH_{J} = J \sum_i (\sigma_{i1}^z\sigma_{i2}^z + \sigma_{i2}^z\sigma_{i3}^z)$ selects two configurations on every column: $\ket{\sigma_{i1}^z, \sigma_{i2}^z, \sigma_{i3}^z} = \ket{ \uparrow \downarrow \uparrow }_i$ and $\ket{ \downarrow \uparrow \downarrow}_i$ with $Z_i=\pm1$, respectively.
Consequently, a system with $N_\mathrm{col}$ columns has $2^{N_\mathrm{col}}$ degenerate ground states when $K=0$. To understand how this degeneracy is lifted by a small $K$, note that the nontrivial effect of $\tilde{\HH}_K$ on two neighboring columns, $i$ and $j$, is to flip all pairs of spins $\{ \sigma_{i\ell}^z, \sigma_{j\ell}^z \}$ that occupy the same layer and are antialigned. 
Hence, $\tilde{\HH}_K$ induces fluctuations to virtual states whenever $i$ and $j$ are in opposite low-energy configurations, $Z_i = -Z_j$. Using second-order perturbation theory, one can show that these processes are responsible for stabilizing the antiferromagnetic order in $Z_i$.

The FSDW phase breaks $G$ down to $H_1=\UonecrossUone$ and has $\Delta n_A=0$. One possibility after spontaneous symmetry breaking is for $\vec{m}$ to lie in the $m_3$-$m_8$ plane. In this case, Eq.~\eqref{eq:nil FSDW} applies and only the outer layers develop staggered local magnetizations, so that $\expval{\sigma_{i1}^z} = - \expval{\sigma_{i3}^z}$ while $\expval{\sigma_{i2}^z}=0$. Thus, the local order parameter is simply the staggered magnetization $\expval{\sigma_{i3}^z - \sigma_{i1}^z}$ as shown in Fig.~\ref{fig:localorder_phases}.
More generally, a FSDW state does not need to have $\vec{m}$ restricted to the $m_3$-$m_8$ plane, but can develop nonzero $m_4$ and $m_5$ components. Correspondingly, string correlators $C_\gamma^4$ and $C_\gamma^5$ will carry the nonlocal part of the order.

The mixed phase has the same broken symmetry $H_1$ as the FSDW, but also develops $\Delta n_A \neq 0$. Its local content is therefore the combination of the CDW order in Eq.~\eqref{eq:map_cdw} and a staggered outer-layer polarization, with $\expval{\sigma_{i1}^z}$, $\expval{\sigma_{i2}^z}$, and $\expval{\sigma_{i3}^z}$ all generically nonzero in the diagonal representative as shown in Fig. \ref{fig:localorder_phases}. As in the FSDW phase, generic $H_1$ states may also develop nondiagonal components in the $\{1,3\}$ sector, so that their nonlocal content is described by $C_\gamma^4$ and $C_\gamma^5$.

%%%%%%%%%%%%%%%%%%%%%%%%%%%%%%%%%%%%%%%%%%%%%%%%%%%%%%%%%%%
%%%%%%%%%%%%%%%%%%%%%%%%%%%%%%%%%%%%%%%%%%%%%%%%%%%%%%%%%%%

\subsection{Phases at the SU(3)-symmetric point}

At $J'=J$, the fixed-flux Hamiltonian acquires an enhanced $G=\mathrm{SU}(3)$ symmetry.  As discussed in Sec. \ref{subsec:mftn3_phases+pd}, it is useful to classify the resulting ordered states by diagonalizing $M=\vec{m}\cdot\bm{\lambda}$, since the eigenvalue structure of $M$ determines the residual symmetry $H$. If $M$ has two degenerate eigenvalues, the ordered state has $H_2=\mathrm{SU(2)}\times \mathrm{U}(1)$. 
If $M$ has no degenerate eigenvalues, the residual symmetry is $H_1=\mathrm{U(1)}\times \mathrm{U(1)}$. The local content of the order is obtained whenever $\vec{m}$ lies in the $m_3$-$m_8$ plane. In that case, the ordered state is described by staggered diagonal combinations of $S_i^3$ and $S_i^8$, and therefore by staggered linear combinations of $\sigma_{i1}^z$, $\sigma_{i2}^z$, and $\sigma_{i3}^z$ through Eq.~\eqref{eq:map_local_diag}. For phase I (Fig. \ref{fig:phases}), $\Delta n_A \neq 0$, this local order implies a staggered total magnetization in Eq.~\eqref{eq:map_cdw}. A set of possible corresponding local magnetization configurations for phases I ($H_2$) and II ($H_1$) phases are those already described in the previous section and depicted in Fig.~\ref{fig:localorder_phases}.
When $M$ has components along $\lambda^a$ with $a\ne 3,8$, the corresponding part of the order is not local in the microscopic variables and is instead encoded by the string correlators $C_\gamma^a$ in Eq.~\eqref{eq:Ca def}. Thus, both $H_1$ and $H_2$ phases on the isotropic line may contain purely local order, purely nonlocal string order, or a coexistence of both, depending on the decomposition of $\vec{m}$ into diagonal and nondiagonal components in the original layer basis.

Throughout the phase diagram, while the spin sector exhibits the local orders discussed above, the orbital sector remains in a liquid state supporting the nonlocal string order. Phases that display such coexistence of Landau-type order parameters and nonlocal topological order are said to show ``magnetic fragmentation'' \cite{vijayvargia23,bartlett14,petit16}.

%%%%%%%%%%%%%%%%%%%%%%%%%%%%%%%%%%%%%%%%%%%%%%%%%%%%%%%%%%%
%%%%%%%%%%%%%%%%%%%%%%%%%%%%%%%%%%%%%%%%%%%%%%%%%%%%%%%%%%%
%%%%%%%%%%%%%%%%%%%%%%%%%%%%%%%%%%%%%%%%%%%%%%%%%%%%%%%%%%%

\section{Summary and outlook}
\label{sec:summary}

In summary, we constructed a family of layered Hamiltonians where each member, composed of $N$ identical layers of a two-dimensional quantum spin-orbital model, realizes an emergent SU($N$) symmetry in the limit of equal all-to-all interlayer interactions. We showed how, at low energies, these models can be mapped to $N$-component Hubbard models on a $\pi$-flux square lattice. Furthermore, we invoked symmetry-based arguments to predict that the models with $N>2$ harbor a rich $T=0$ phase diagram in the proximity of their SU($N$)-symmetric limit, and then supported this claim by presenting an explicit mean-field calculation for $N=3$. Finally, we elucidated the connection between the ground states of the $N=3$ system and different forms of magnetic fragmentation.

Interesting directions for future work include the exploration not only of unconventional phases, but also of quantum critical properties that arise in spin-orbital models with different values of $N$.
In view of their correspondence to effective Hubbard models of relativistic fermions (see Sec.~\ref{subsec:mapping}), our layered Hamiltonians may -- at least in their SU($N$)-symmetric limits -- realize a slew of quantum phase transitions in different Gross-Neveu* universality classes \cite{seifert20,fornoville25}. The latter are fractionalized variants of the Gross-Neveu universality classes, which were first discussed in the context of high-energy physics \cite{nambu61,gross74}, but later identified as relevant to graphene \cite{herbut06,herbut09a,herbut09b,boyack21} and related moiré systems \cite{ma24,biedermann25,huang25}.
In particular, the link to Gross-Neveu* criticality is guaranteed to apply in the $N=2$ case, where tuning $J/K$ drives a transition between the spin-orbital liquid and a Néel antiferromagnet \cite{vijayvargia23} that falls within the Gross-Neveu-Heisenberg* universality class \cite{ixert14,toldin15}.

%\pc{Include something about CP$^N$ skyrmions? \cite{williams25}}

Finally, we note that one can effectively dope the spin-orbital Hamiltonians considered above with the addition of a chemical potential term
\begin{equation}
    \HH_\mu = \frac{\mu}{2} \sum_{i\ell} \Gamma_{i\ell}^5 = -\mu \sum_{i\ell} \left( f_{i\ell}^\dagger f_{i\ell} - \frac{1}{2} \right).
\end{equation}
Since Lieb's theorem does not apply away from half filling \cite{lieb94}, this may provide a useful connection to the physics of $N$-component Hubbard models in different flux sectors. It may also reproduce different phases discussed in past studies of the large-$N$ limit of SU($N$) Hubbard-Heisenberg models \cite{affleck88,marston89}, including a chiral spin liquid expected to emerge at strong coupling for average filling of one fermion per site \cite{hermele09}.

%%%%%%%%%%%%%%%%%%%%%%%%%%%%%%%%%%%%%%%%%%%%%%%%%%%%%%%%%%%
%%%%%%%%%%%%%%%%%%%%%%%%%%%%%%%%%%%%%%%%%%%%%%%%%%%%%%%%%%%
%%%%%%%%%%%%%%%%%%%%%%%%%%%%%%%%%%%%%%%%%%%%%%%%%%%%%%%%%%%

\begin{acknowledgments}

We thank Lukas Janssen and  Tanmay Vachaspati for illuminating discussions. PMC is also grateful to  Matthias Vojta for a collaboration on a previous related project. PMC was supported by the U.S. Department of Energy, Office of Science, Office of Basic Energy Sciences, Material Sciences and Engineering Division under Award Number DE-SC0025247. AV and OE acknowledge support from NSF Award No.
DMR-2234352. The authors acknowledge Research Computing at Arizona State University for providing HPC resources \cite{Sol} that contributed to the results reported in this paper.

\end{acknowledgments}

%%%%%%%%%%%%%%%%%%%%%%%%%%%%%%%%%%%%%%%%%%%%%%%%%%%%%%%%%%%
%%%%%%%%%%%%%%%%%%%%%%%%%%%%%%%%%%%%%%%%%%%%%%%%%%%%%%%%%%%
%%%%%%%%%%%%%%%%%%%%%%%%%%%%%%%%%%%%%%%%%%%%%%%%%%%%%%%%%%%

\appendix

%%%%%%%%%%%%%%%%%%%%%%%%%%%%%%%%%%%%%%%%%%%%%%%%%%%%%%%%%%%
%%%%%%%%%%%%%%%%%%%%%%%%%%%%%%%%%%%%%%%%%%%%%%%%%%%%%%%%%%%
%%%%%%%%%%%%%%%%%%%%%%%%%%%%%%%%%%%%%%%%%%%%%%%%%%%%%%%%%%%

\section{Mean-field decoupling and\\$\mathrm{SU}(N)$ basis transformation}
\label{appendix:mft}

In this appendix, we outline the derivation of the mean-field Hamiltonian in Sec.~\ref{sec:mft_general} and present the basis transformation relating the original of mean fields to the average filling per site, $n_i$, and the ``magnetization'' $\vec{m}_i$ defined in Eq.~\eqref{eq:mia}.

The mean-field form \eqref{eq:HJ_MF_O_form} of the interaction term $\tilde{\HH}_J$ is derived by decoupling Eq.~\eqref{eq:HhubbJ} in the density-density (Hartree) and particle-hole (Fock) channels. Thus, we keep the full local $N\times N$ matrix
\begin{equation}
    O_i^{\ell\ell'}\equiv \big\langle f^\dagger_{i\ell} f_{i\ell'}\big\rangle
\label{eq:Oi_def_app}
\end{equation}
as a mean field. For each unordered pair $\ell<\ell'$, the quartic operator admits the standard Hartree-Fock decomposition
\begin{align}
    f^\dagger_{i\ell}f^\dagger_{i\ell'}f_{i\ell'}f_{i\ell}
    &\approx
    O_{i}^{\ell'\ell'}\,\hat{n}_{i\ell} + O_{i}^{\ell\ell}\,\hat n_{i\ell'}-O_{i}^{\ell\ell} O_{i}^{\ell'\ell'}
    \notag \\
    &-\Bigl(O_i^{\ell\ell'} f^\dagger_{i\ell'}f_{i\ell}+O_i^{\ell'\ell} f^\dagger_{i\ell}f_{i\ell'}-|O_i^{\ell\ell'}|^2\Bigr).
    \label{eq:HF_decouple_pair_app}
\end{align}
Equation~\eqref{eq:HJ_MF_O_form} is then obtained by writing Eq.~\eqref{eq:HhubbJ} in normal order, using Eq.~\eqref{eq:HF_decouple_pair_app} and the property $J_{\ell\ell'} = J_{\ell'\ell}$, and dropping an additive constant independent of the mean fields.

As discussed in the main text, the symmetry-breaking mechanism can be made more explicit by expressing the mean-field Hamiltonian in terms of the order parameters $\vec{m}_i$. To do so, we make use of the fact that $O_i$ is a $N\times N$ Hermitian matrix, and therefore can be expanded in a basis formed by the $N\times N$ identity matrix $\mathds{1}$ and (complex conjugates of the) the generators $\eta^a$ of $\mathrm{SU}(N)$: $O_i = c_0 \mathds{1} + c_a (\eta^a)^*$. The coefficients of the expansion can be determined by using the property $\Tr(\eta^a\eta^b)=2\delta^{ab}$:
\begin{equation}
    O_i^{\ell\ell'}
    =
    \frac{n_i}{N}\,\delta_{\ell\ell'}
    +\sum_{a=1}^{N^2-1} m_{ia}\,\eta^a_{\ell'\ell}.
    \label{eq:O_decomp_app}
\end{equation}
The diagonal elements $\ell = \ell'$ are nothing but the flavor-resolved densities
\begin{equation}
    n_{i\ell} = \frac{n_i}{N} + \vec m_i \cdot \bm{\eta}_{\ell\ell}.
    \label{eq:Oill_diag_transformation}
\end{equation}
Note that this relation reduces to Eq.~\eqref{eq:nil N=3} for $N=3$. 

Using Eq.~\eqref{eq:Oill_diag_transformation} to rewrite the flavor-diagonal elements $O_i^{\ell\ell}$ in Eq.~\eqref{eq:HJ_MF_O_form} in terms of $n_i$ and $\vec{m}_i$, we obtain
\begin{align}
    \tilde{\HH}_{J,\mathrm{MF}} &=
    4 \sum_{i,\ell} \sum_{\ell' \ne \ell} J_{\ell \ell'} \left[
    \frac{1}{2} \abs{ M_i^{\ell'\ell} }^2 \right.
    \notag \\    & 
    -\frac{1}{2} \left( \frac{n_i}{N} + M_i^{\ell\ell} \right)
    \left( \frac{n_i}{N} + M_i^{\ell'\ell'} \right)
    \notag \\
    & \left. +\left(
    \frac{n_{i}}{N} + M_i^{\ell'\ell'} - \frac{1}{2}
    \right) \hat{n}_{i\ell}
    - f_{i\ell'}^\dagger M_i^{\ell'\ell} f_{i\ell}
    \right],
    \label{eq:HJ_MF_m_form_app}
\end{align}
where $M_i = \vec{m}_i \cdot \bm{\eta}$ is the traceless part of $O_i$. One can show that this expression simplifies to Eq.~\eqref{eq:HJ_MF_sun} when $J_{\ell\ell'} = J$ by using the property $\Tr (\eta^a) = 0$.

%%%%%%%%%%%%%%%%%%%%%%%%%%%%%%%%%%%%%%%%%%%%%%%%%%%%%%%%%%%
%%%%%%%%%%%%%%%%%%%%%%%%%%%%%%%%%%%%%%%%%%%%%%%%%%%%%%%%%%%

\subsection{Generating symmetry-equivalent mean-field states}

Given a set of mean-field parameters $\{ n_i, \vec{m}_i \}$, one can generate a different set $\{ n_i', \vec{m}_i' \}$ by applying a transformation $U \in \mathrm{U}(N)$ to the matrices $O_i$ in Eq.~\eqref{eq:O_decomp_app}. Concretely,
\begin{equation}
    O_i' = U^\dagger O_i U \equiv 
    \frac{n_i'}{N} \mathds{1} + \vec{m}_i' \cdot \bm{\eta}^{\top},
\end{equation}
where $\top$ is used to indicate the transpose of a matrix. If $U$ further belongs to the symmetry group $G$ of the Hamiltonian, then the states described by $\{ n_i, \vec{m}_i \}$ and $\{ n_i', \vec{m}_i' \}$ are guaranteed to be degenerate. The new parameters can be computed directly via
\begin{align}
    n_i' &= \Tr (O_i') = \Tr (O_i) = n_i,
    \\
    \vec{m}_i' &= \frac{1}{2} \Tr \left( O_i' \bm{\eta}^\top \right)
    = \frac{1}{2} \Tr \left( U^\dagger M_i^\top U \bm{\eta}^\top \right),
\end{align}
with $M_i = \vec{m}_i \cdot \bm{\eta}$.

%%%%%%%%%%%%%%%%%%%%%%%%%%%%%%%%%%%%%%%%%%%%%%%%%%%%%%%%%%%
%%%%%%%%%%%%%%%%%%%%%%%%%%%%%%%%%%%%%%%%%%%%%%%%%%%%%%%%%%%
%%%%%%%%%%%%%%%%%%%%%%%%%%%%%%%%%%%%%%%%%%%%%%%%%%%%%%%%%%%

\section{Symmetries of the mean-field solutions of the $\pi$-flux SU(3) Hubbard model}
\label{appendix:su3 proof}

This appendix presents a derivation of the conditions under which the mean-field Hamiltonian \eqref{eq:HJ_MF_sun} with $N=3$ and $\vec{m}_i \propto \vec{m}$ has an enhanced $\mathrm{SU(2)} \times \mathrm{U(1)}$ symmetry. 

As explained in Sec.~\ref{sec:mft_general}, the symmetry group $H$ 
of a mean-field Hamiltonian $\tilde{\HH}$ with $\vec{m} \ne \vec{0}$ is determined by finding the set of matrices $T$ that commute with $M = \vec{m} \cdot \bm{\lambda}$.
For a general order parameter $\vec{m}$, the solution to this problem starts with the diagonalization of $M$ via a unitary transformation $U$. If $\{ \mu_1, \mu_2, \mu_3 \}$ denote the real eigenvalues of $M$, then
\begin{equation}
    M' = U^\dagger M U = 
    \begin{pmatrix}
        \mu_1 & 0 & 0 \\
        0 & \mu_2 & 0 \\
        0 & 0 & \mu_3
    \end{pmatrix}.
    \label{eq:Mprime}
\end{equation} 
Since $\Tr M' = \Tr M = 0$, we have $\mu_1 + \mu_2 + \mu_3 = 0$; furthermore, Eq.~\eqref{eq:Mprime} can be expanded as a linear combination of the diagonal Gell-Mann matrices: $M' = m_3' \lambda^3 + m_8' \lambda_8$.

In this diagonal basis, the condition $\comm{M'}{T'} = 0$ yields
\begin{equation}
    (\mu_i - \mu_j) T_{ij}' = 0,
    \label{eq:Tprime condition}
\end{equation}
i.e., every nondegenerate pair of eigenvalues $(\mu_i, \mu_j)$ places a constraint on the set of solutions $T'$. Thus, we separate our analysis into two scenarios\footnote{Note that, since $M$ is traceless, it can only have a fully degenerate spectrum if $\mu_1 = \mu_2 = \mu_3 = 0$. We do not consider this trivial case because it implies that $\vec{m}=\vec{0}$, and therefore corresponds to a symmetric (symmetry-unbroken) phase.}: (i) The spectrum of $M$ is nondegenerate or (ii) has a partial degeneracy, $\mu_1 = \mu_2 \ne \mu_3$.

In scenario (i), Eq.~\eqref{eq:Tprime condition} implies that the only solutions are diagonal matrices $T' = t_3' \lambda^3 +t_8' \lambda^8$. Given that SU(3) has rank 2, it is always possible obtain two mutually commuting matrices of this form. Each of them can be transformed back to the original basis, $T = U T' U^\dagger$, where they will act as generators of disjoint U(1) symmetries. Consequently, the mean-field Hamiltonian will be invariant under $H_1 = \mathrm{U(1)} \times \mathrm{U(1)}$ transformations.

In scenario (ii), the situation changes because Eq.~\eqref{eq:Tprime condition} does not restrict the values of the nondiagonal elements $T'_{12} = (T'_{21})^*$. Hence, a general solution takes the form $T' = \sum_{a=1}^3 t_a' \lambda^a + t_8' \lambda^8$, which results in an invariant subgroup $H_2 = \mathrm{SU(2)} \times \mathrm{U(1)}$.

Now consider
\begin{align}
    E_{\bm{\alpha}_1} &= \frac{1}{2} (\lambda^1 + \ii \lambda^2)
    = \begin{pmatrix}
        0 & 1 & 0 \\
        0 & 0 & 0 \\
        0 & 0 & 0
    \end{pmatrix}
    = \ket{1}\bra{2},
    \\
    E_{\bm{\alpha}_2} &= \frac{1}{2} (\lambda^6 + \ii \lambda^7)
    = \begin{pmatrix}
        0 & 0 & 0 \\
        0 & 0 & 1 \\
        0 & 0 & 0
    \end{pmatrix}
    = \ket{2}\bra{3},
    \\
    E_{\bm{\alpha}_3} &= \frac{1}{2} (\lambda^4 + \ii \lambda^5)
    = \begin{pmatrix}
        0 & 0 & 1 \\
        0 & 0 & 0 \\
        0 & 0 & 0
    \end{pmatrix}
    = \ket{3}\bra{1}.
\end{align}
Given a diagonal matrix $M'$ from Eq.~\eqref{eq:Mprime}, we have
\begin{align}
    \comm{M'}{E_{\bm{\alpha}_n}} = \mu_n - \mu_{n+1},
    \label{eq:Mprime_comm1}
\end{align}
where we have defined $\mu_4 = \mu_1$ for brevity. At the same time, it is straightforward to verify that
\begin{align}
    \comm{\lambda^3}{E_{\bm{\alpha}_n}} &= 2\alpha_n^{(1)} E_{\bm{\alpha}_n},
    \\
    \comm{\lambda^8}{E_{\bm{\alpha}_n}} &= 2\alpha_n^{(2)} E_{\bm{\alpha}_n},
    \label{eq:Ealpha_comm}
\end{align}
with $\alpha_n^{(i)}$ being the $i$-th component of the root vectors $\bm{\alpha}_1 = (1,0)$, $\bm{\alpha}_2 = (-1,\sqrt{3})/2$, and $\bm{\alpha}_3 = (1,\sqrt{3})/2$. Thus, if we write $M' = m_3' \lambda^3 + m_8' \lambda^8$ and let $\vec{m}' = (m_3,m_8)$, we find
\begin{equation}
    \comm{M'}{E_{\bm{\alpha}_n}} = 2 \vec{m}' \cdot \bm{\alpha}_n.
    \label{eq:Mprime_comm2}
\end{equation}
By comparing Eqs.~\eqref{eq:Mprime_comm1} and \eqref{eq:Mprime_comm2}, we arrive at the same result stated in the main text: Scenario (ii) only arises if $\vec{m}'$ is perpendicular to one of the three roots $\bm{\alpha}_n$ of the SU(3) algebra.

%%%%%%%%%%%%%%%%%%%%%%%%%%%%%%%%%%%%%%%%%%%%%%%%%%%%%%%%%%%
%%%%%%%%%%%%%%%%%%%%%%%%%%%%%%%%%%%%%%%%%%%%%%%%%%%%%%%%%%%
%%%%%%%%%%%%%%%%%%%%%%%%%%%%%%%%%%%%%%%%%%%%%%%%%%%%%%%%%%%

\section{Proof of the emergence of long-range string order}
\label{appendix:string}

This appendix presents a proof of the statement leading to Eqs.~\eqref{eq:Ca def} and \eqref{eq:Bdef} in Sec.~\ref{sec:mapback}. As a reminder, we are interested in showing that, if the ground state $\ket{\psi_u}$ of the Hamiltonian $\tilde{\HH}$ within a fixed gauge sector $\left\{ u \right\}$ displays long-range, SU($N$) symmetry-breaking order characterized by
\begin{equation}
    m_a^u = \frac{1}{N_s} \sum_i (-1)^i \expval{S_i^a}{\psi_u} \ne 0
\end{equation}
in the thermodynamic limit $N_s \to \infty$, then the corresponding physical ground state $\ket{\psi} = P \ket{\psi_u}$ will display nontrivial string correlations.

To prove this claim, it will be convenient to use the factorization \cite{yao09}
\begin{equation}
	P = P' \; \frac{(1+D)}{2}
	\label{eq:P factorization}
\end{equation}
of the projection operator in Eq.~\eqref{eq:projector}. Here, $D = \prod_{i\ell} D_{i\ell}$ implements a gauge transformation \eqref{eq:majorana gauge} on every site and $P'$ constructs an equal-weight superposition of all equivalent gauge-field configurations. Because $D^2 = 1$ and  $\comm{\HH}{D} = 0$, $\ket{\psi_u}$ is an eigenstate of $D$ with eigenvalue $\pm1$. If we take this into account and use the fact that $D$ does not change any of the gauge fields (it flips each $u_{\ell,ij}$ twice), it becomes clear that the role of the factor $(1+D)$ in Eq.~\eqref{eq:P factorization} is to project out unphysical states.

Now suppose that $\ket{\psi_u}$ and $\ket{\psi_{\tilde{u}}}$ are two states related by a gauge transformation $\tilde{D}$, i.e., $D\ket{\psi_u} = \ket{\psi_u}$, $D \ket{\psi_{\tilde{u}}} = \ket{\psi_{\tilde{u}}}$, and $\ket{\psi_{\tilde{u}}} = \tilde{D} \ket{\psi_u}$. Then, if $O$ is a gauge-invariant observable obeying $\comm{O}{D_{i\ell}} = 0$ for every site $(i\ell)$ and $\comm{O}{\hat{u}_{\ell,ij}} = 0$ for every pair of sites $(i\ell,j\ell)$, we have
\begin{align}
	\matrixel{\psi_u}{O}{\psi_{\tilde{u}}} &= \delta_{u, \tilde{u}} \expval{O}{\psi_u},
	\label{eq:prop1}
    \\
	\expval{O}{\psi_{\tilde{u}}} 
    &= \expval{\tilde{D} O \tilde{D}}{\psi_u} 
    = \expval{O}{\psi_u}.
	\label{eq:prop2}
\end{align}
Using these two properties, we find that
\begin{align}
	C_\gamma^a \left(\vec{r}_i, \vec{r}_j\right) 
	&\propto \sum_{u, \tilde{u}} \matrixel{\psi_u}{S_{i}^a \hat{B}^a_\gamma \left(\vec{r}_i, \vec{r}_j\right) S_{j}^a}{\psi_{\tilde{u}}}
	\notag \\
	&= N_\mathrm{gc} B^a_\gamma \left(\vec{r}_i, \vec{r}_j\right) \expval{S_{i}^a S_{j}^a}{\psi_u}.
	\label{eq:Ca proof}
\end{align}
where $N_\mathrm{gc}$ denotes the number of equivalent gauge-field configurations and $B^a_\gamma \left(\vec{r}_i, \vec{r}_j\right) = \pm 1$ is an integer obtained after replacing the $\hat{u}_{\ell,ij}^\alpha$ operators in Eq.~\eqref{eq:Bdef} by the corresponding eigenvalues in $\left\{u \right\}$.
Therefore, $C_\gamma^a \left(\vec{r}_i, \vec{r}_j\right)$ is proportional to $\expval{S_{i}^a S_{j}^a}{\psi_u}$ and will be nonzero for every $(ij)$ as long as the unphysical state $\ket{\psi_u}$ develops long-range order.

%%%%%%%%%%%%%%%%%%%%%%%%%%%%%%%%%%%%%%%%%%%%%%%%%%%%%%%%%%%
%%%%%%%%%%%%%%%%%%%%%%%%%%%%%%%%%%%%%%%%%%%%%%%%%%%%%%%%%%%
%%%%%%%%%%%%%%%%%%%%%%%%%%%%%%%%%%%%%%%%%%%%%%%%%%%%%%%%%%%

\bibliography{layered_somodels}

%%%%%%%%%%%%%%%%%%%%%%%%%%%%%%%%%%%%%%%%%%%%%%%%%%%%%%%%%%%

\end{document}